\documentclass[a4paper,twoside,12pt]{article}

\usepackage{xspace}
\usepackage{graphicx}
\usepackage{subfigure}
\usepackage{placeins}
\usepackage[top=1in, bottom=1.25in, left=1.25in, right=1.25in]{geometry}
\usepackage{hyperref}

\newcommand*{\antibar}[1]{\ensuremath{#1\bar{#1}}\xspace}

\newcommand*{\ttbar}{\antibar{t}}
\newcommand*{\bbbar}{\antibar{b}}
\newcommand*{\ppbar}{\antibar{p}}
\newcommand*{\qqbar}{\antibar{q}}
\newcommand*{\ttW}{\ensuremath{\ttbar W}\xspace}
\newcommand*{\ttZ}{\ensuremath{\ttbar Z}\xspace}
\newcommand*{\ttgam}{\ensuremath{\ttbar \gamma}\xspace}
\newcommand*{\ttH}{\ensuremath{\ttbar H}\xspace}
\newcommand*{\TeV}{\ifmmode {\mathrm{\ Te\kern -0.1em V}}\else
                   \textrm{Te\kern -0.1em V}\fi}%
\newcommand*{\GeV}{\ifmmode {\mathrm{\ Ge\kern -0.1em V}}\else
                   \textrm{Ge\kern -0.1em V}\fi}%
\newcommand*{\pt}{\ensuremath{p_{\rm T}}\xspace}
\newcommand{\subfig}[3]{\raisebox{#3pt}{\subfigure{\includegraphics[width=#2\textwidth]{#1}}}}
\newcommand{\stt}{\ensuremath{\sigma_{\ttbar}}}
\newcommand{\stat}{\ensuremath{\:\textrm{(stat.)}}}
\newcommand{\syst}{\ensuremath{\:\textrm{(syst.)}}}
\newcommand{\lumi}{\ensuremath{\:\textrm{(lumi.)}}}
\newcommand{\pb}{\ensuremath{\:\textrm{pb}}}

\newcommand*{\Vtb}{\ensuremath{\vert V_{tb} \vert}\xspace}
\newcommand*{\ipb}{\ensuremath{\textrm{pb}^{-1}}\xspace}

\newcommand*{\MET}{\ensuremath{E_{\mathrm{T}}^{\mathrm{miss}}}\xspace}
\newcommand*{\RHF}{\ensuremath{R_{\mathrm{HF}}}\xspace}
\renewcommand*{\to}{\ensuremath{\rightarrow}\xspace}
\newcommand{\coll}[1]{#1 Collaboration}
\newcommand{\conf}[3]{ATLAS-CONF-#1-#2, http://inspirehep.net/record/#3}
\newcommand{\pas}[3]{CMS PAS TOP-#1-#2, http://inspirehep.net/record/#3}
\newcommand{\arxiv}[2]{arXiv:#1 [hep-#2]}
\newcommand{\subm}[1]{submitted to #1}
\newcommand{\accept}[1]{accepted by #1}
\newcommand{\PRL}{Phys.\ Rev.\ Lett.}
\newcommand{\PRD}{Phys.\ Rev.\ D}
\newcommand{\EPJC}{Eur.\ Phys.\ J.\ C}
\newcommand{\PLB}{Phys.\ Lett.\ B}

\title{\bf{Top-quark production measurements}}
\author{Markus Cristinziani
\thanks{on behalf of the ATLAS, CDF, CMS, D0 and LHCb Collaborations}
\footnote{Supported by European Research Council grant ERC--CoG--617185}\\
Physikalisches Institut, University of Bonn, Germany\\
E-mail: cristinz@uni-bonn.de}

\begin{document}
\maketitle
\begin{abstract}
{Recent measurements of top-quark production at 
hadron colliders are reviewed. The inclusive top-quark pair production is
determined at four centre-of-mass energies at Tevatron and LHC with
experimental uncertainties that are close to the uncertainties in theoretical calculations at
next-to-next-to-leading order in QCD. Several differential measurements are performed and
compared to simulation. Production of single top quarks is studied in the three
different production channels. Top-quark pair production with neutral and charged vector
bosons has been observed by the LHC experiments. Finally, production of 
additional heavy flavour quark pairs (\bbbar, \ttbar) is studied or searched
for.
}
\end{abstract}

\newpage
\section{Introduction}

Top quarks are the elementary particles with the largest mass and are therefore
subject of intensive study at hadron colliders, the proton--antiproton
collider Tevatron, operated until 2011 at up to $1.96 \TeV$ centre-of-mass
energy, and the Large Hadron Collider (LHC) at CERN, operated at energies of
$7$, $8$ and, more recently, $13$ \TeV.  

From the large number of available results, I have chosen to illustrate the
discussion exclusively with results that have been made public in the last
twelve months by the ATLAS, CDF, CMS, D0 and LHCb Collaborations, effectively
covering the full spectrum of studies of top-quark pair production, inclusive
and differential, single top-quark production as well as associated production
with vector bosons and flavoured quarks.  For an overview of recent
developments in the theory of top quarks see Ref.~\cite{melnikov}. Properties
of top quark are presented separately~\cite{properties}.

\section{Inclusive \ttbar production cross section}

Top-quark pair production at Tevatron is dominated by \qqbar annihilation,
while it is dominated by gluon--gluon fusion at the LHC and therefore
measurements at both colliders are complementary and allow to test 
different aspects of perturbative QCD calculations,
which are now known at full next-to-next-to-leading order (NNLO) including
gluon resummation.

\begin{figure}[!h]\centering
\subfig{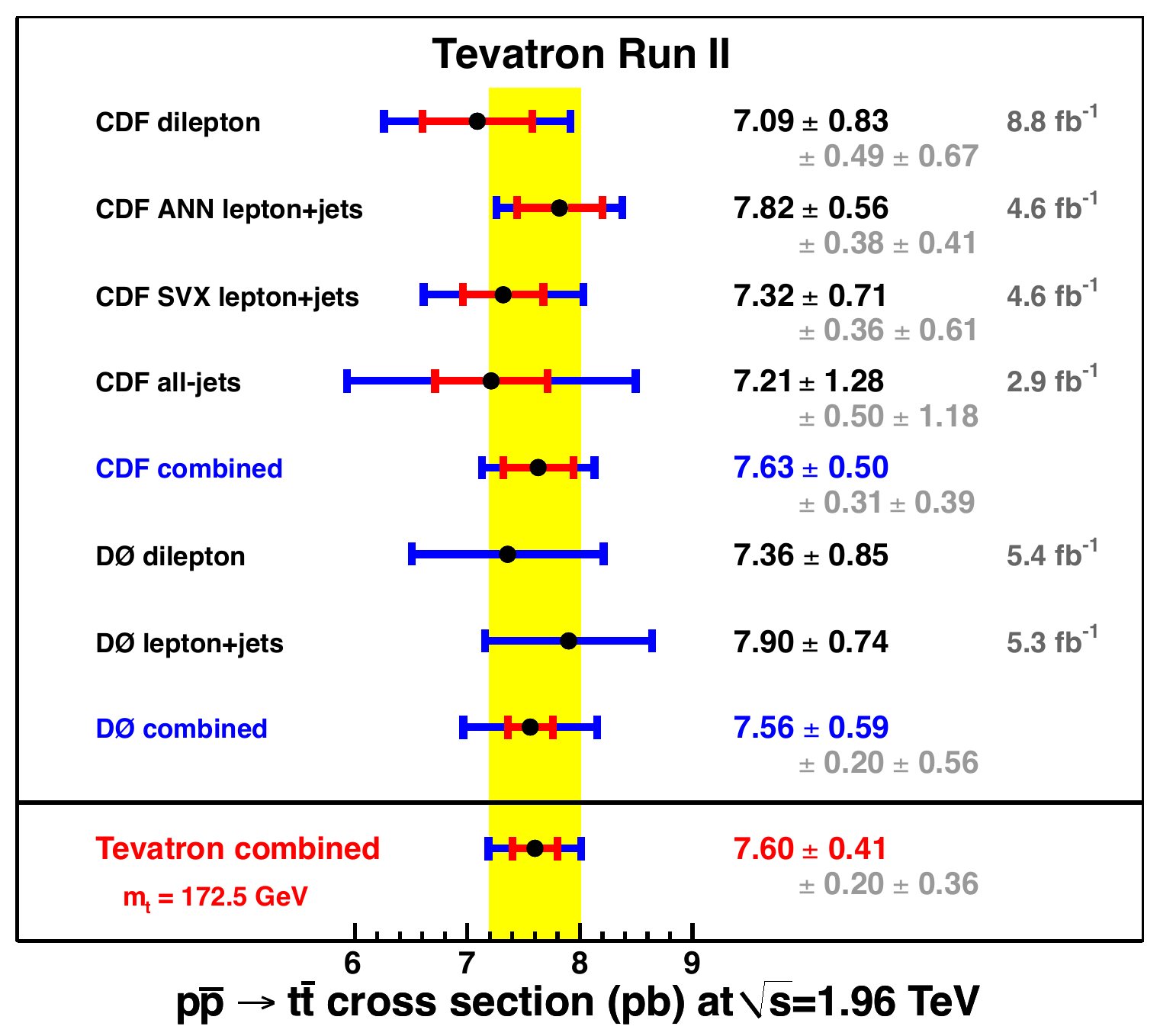}{0.6}{5}\\
\subfig{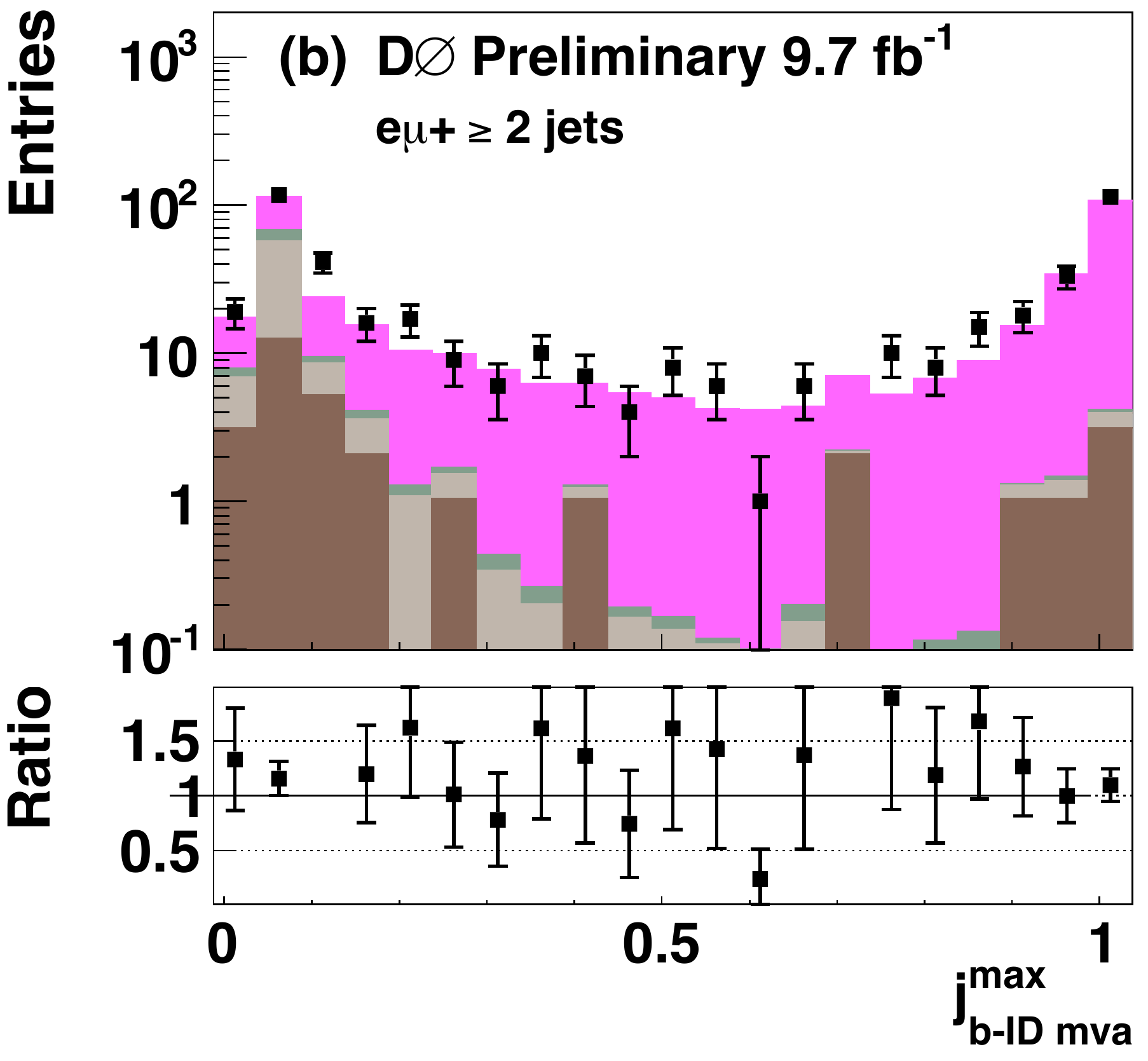}{0.45}{0}
\subfig{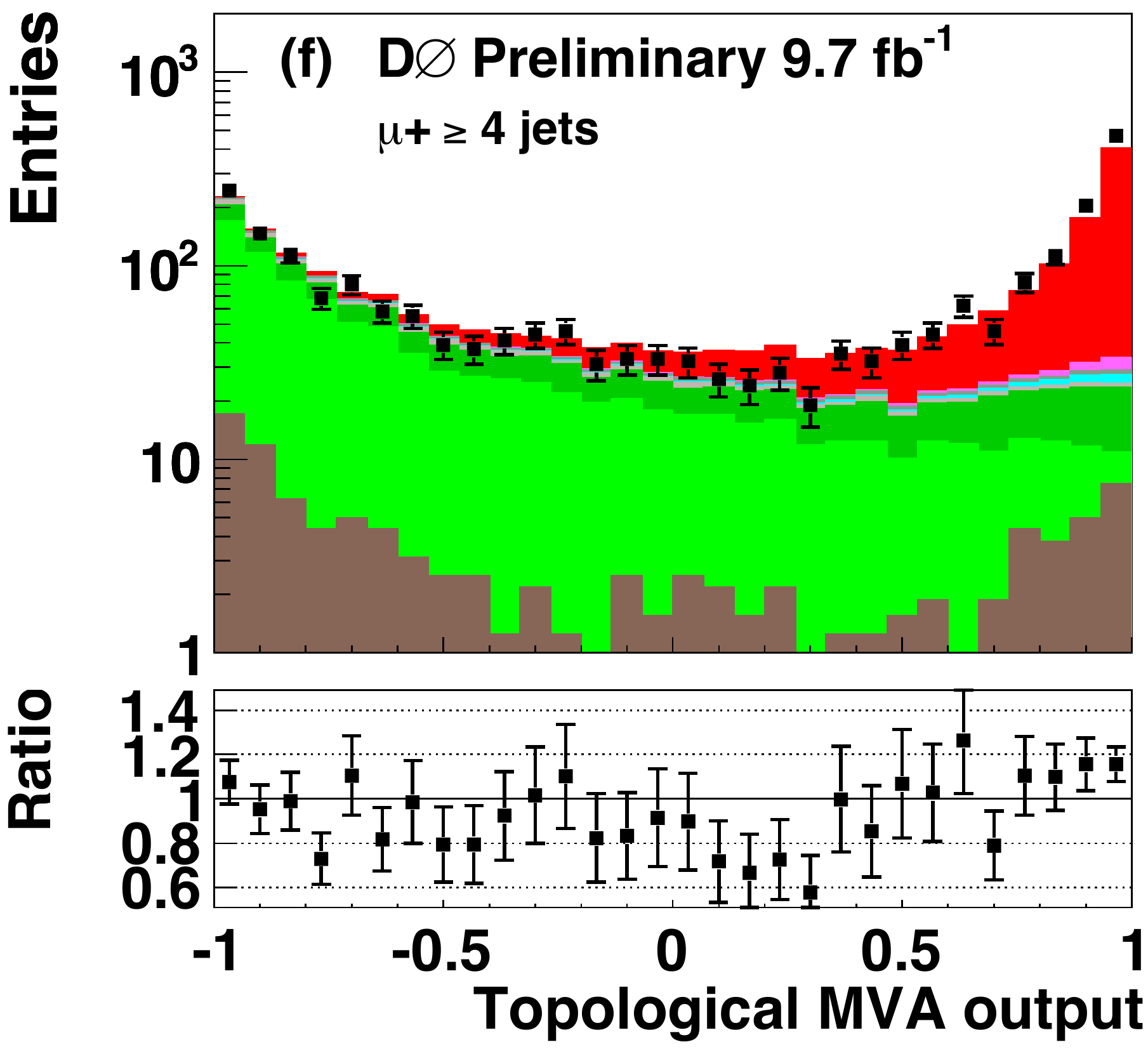}{0.45}{0}
\caption{Summary of \ttbar production cross section measurements at 
Tevatron, showing the most recent combination~\cite{tevcomb} (top), 
the $b$-tagging discriminator
distribution in the D0 dilepton analysis~\cite{d0xs}
(bottom left) and the output variable of the multivariate analysis using topological variables
in the $\ell$+jets channel~\cite{d0xs} (bottom right).}
\label{tt_tev}
\end{figure}

Using the full Tevatron Run II dataset, the D0 Collaboration measures the
inclusive \ttbar cross section~\cite{d0xs} to be $\stt = 7.73 \pm 0.13 \stat
\pm 0.55 \syst \pb$ in final states with one or two leptons (electrons or muons),
exploiting $b$-tagging information.  In the $\ell$+jets channel
additional topological variables are employed (Fig.~\ref{tt_tev}).

\begin{figure}[!h]\centering
\subfig{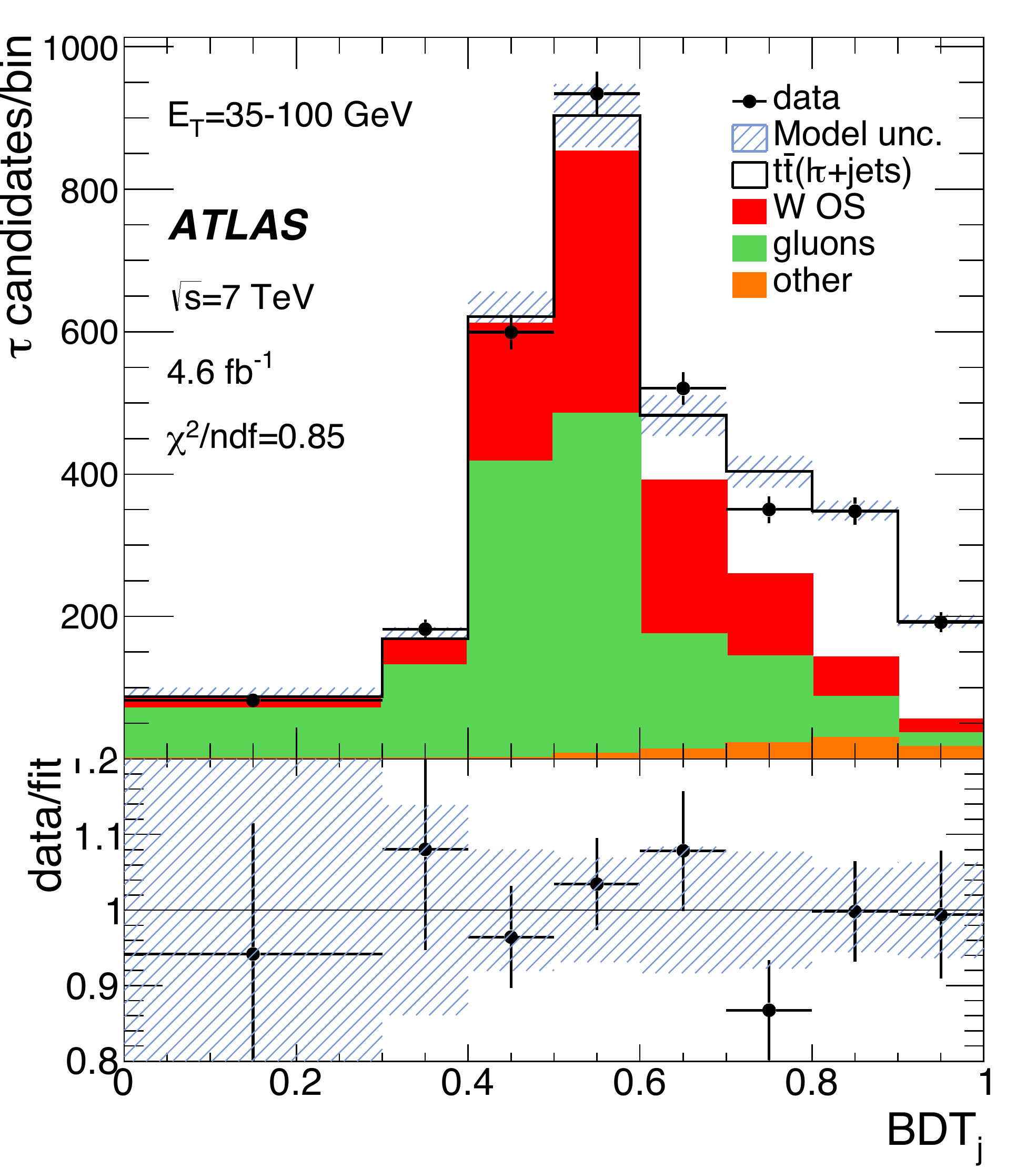}{0.3636}{18}
\hspace{5ex}
\subfig{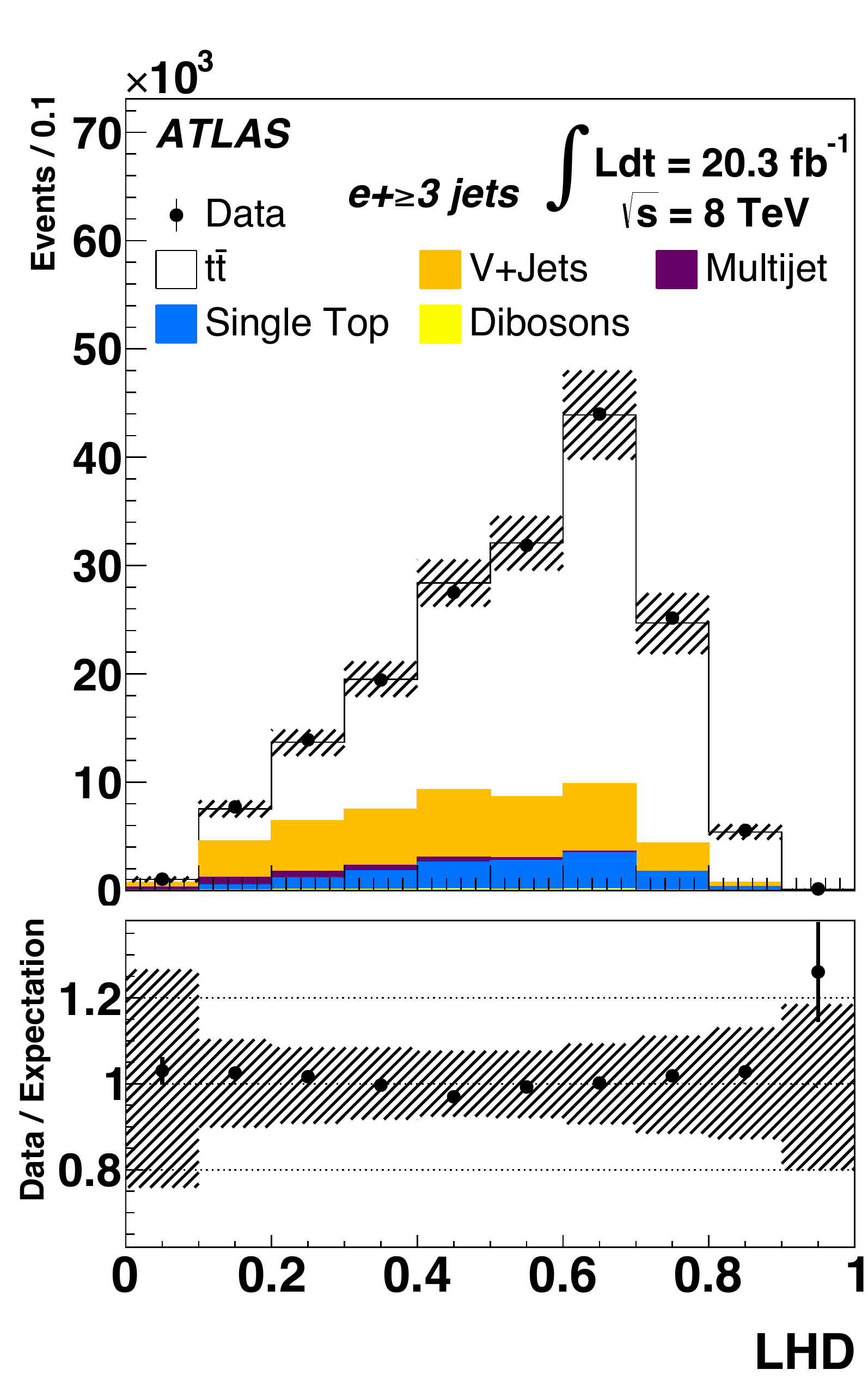}{0.3}{0}
\caption{Left: Fitted distribution of the $\tau$-jet BDT discriminant 
for the $7 \TeV$ \ttbar 
cross-section determination~\cite{alt}. 
Right: Likelihood discriminant in the $e$+jets channel with at least three 
jets~\cite{alj}. \label{tt_7_8}}
\end{figure}

The inclusive \ttbar cross section has been determined in a variety of channels
by ATLAS and CMS at 7 and 8 \TeV.  The $7 \TeV$ dataset is used by ATLAS to
measure the branching ratios of top-quark decays into leptons and
jets~\cite{alt}.  Seven mutually exclusive final states are defined and used in
the analysis, also including two channels with a lepton and a hadronically
decaying $\tau$ lepton (Fig.~\ref{tt_7_8} left).

At $8 \TeV$ a measurement in the single-lepton channel is presented by the ATLAS
Collaboration, using events with at least three jets and at least one
$b$-tagged jet~\cite{alj}.  Systematic uncertainties are smallest, when choosing
only two discriminating input variables, the pseudorapidity of the lepton and
the modified aplanarity, which are combined in a likelihood discriminant
(Fig.~\ref{tt_7_8} right).

\newpage

\begin{figure}[!h]\centering
\subfig{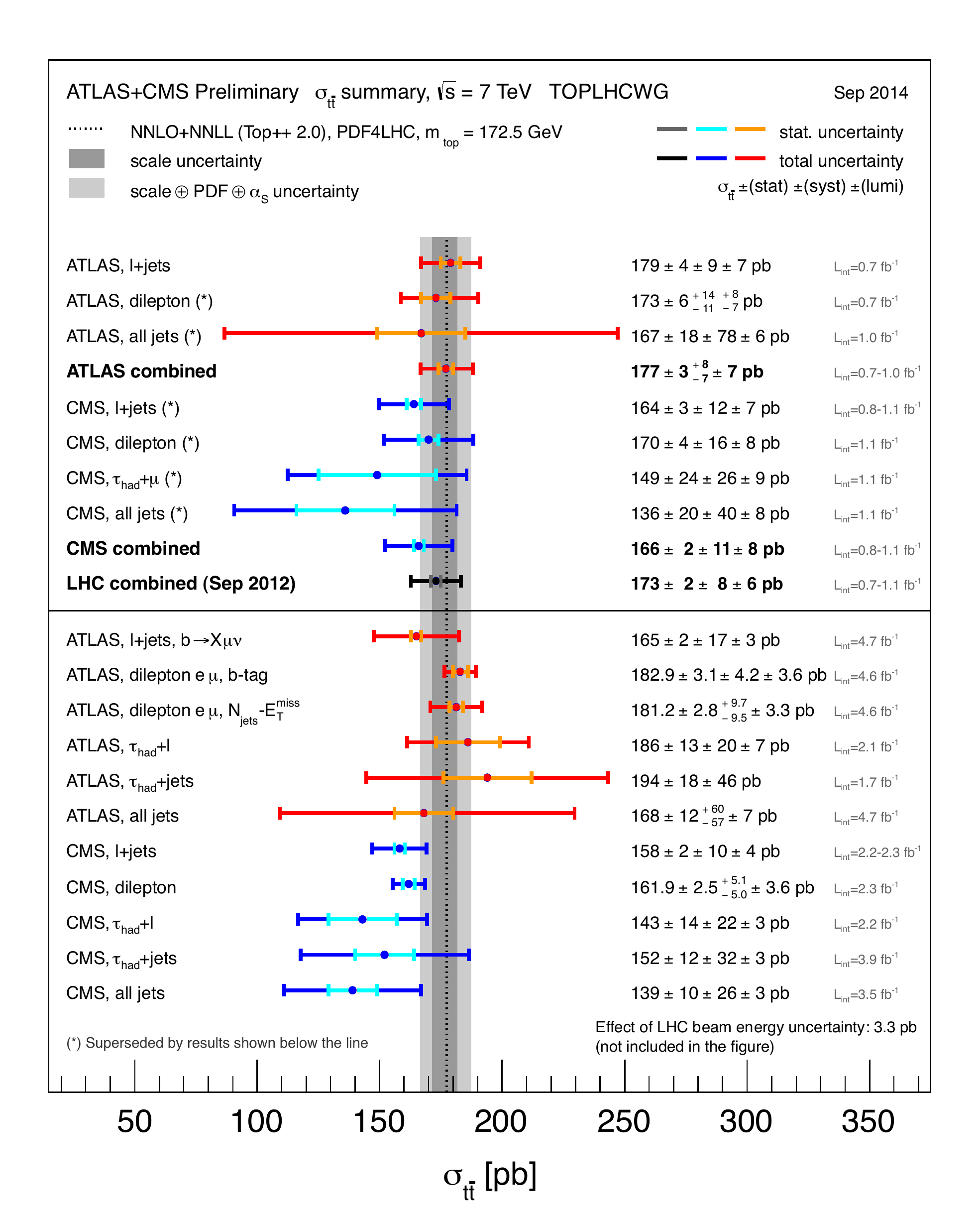}{0.49}{0}
\subfig{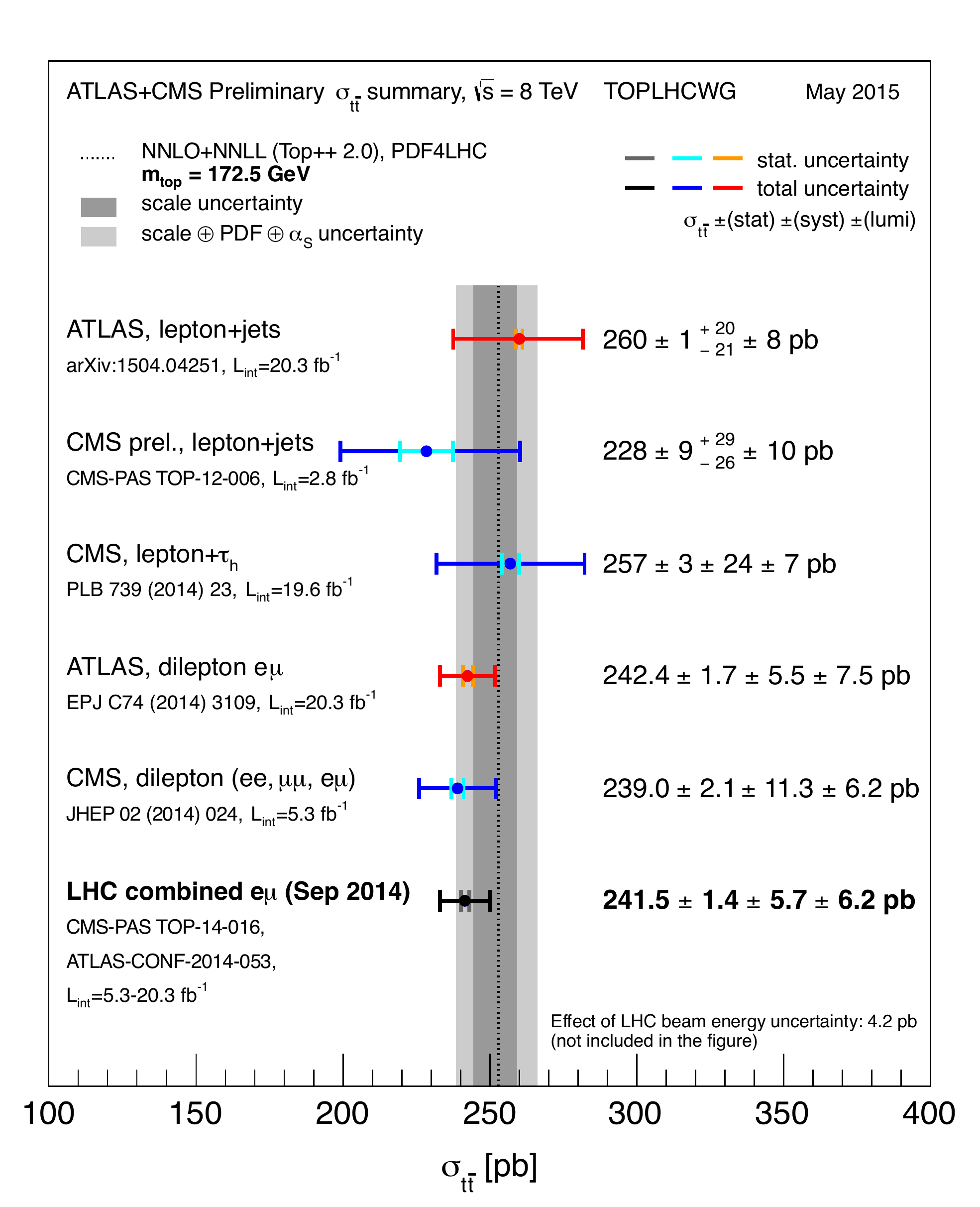}{0.49}{0}
\subfig{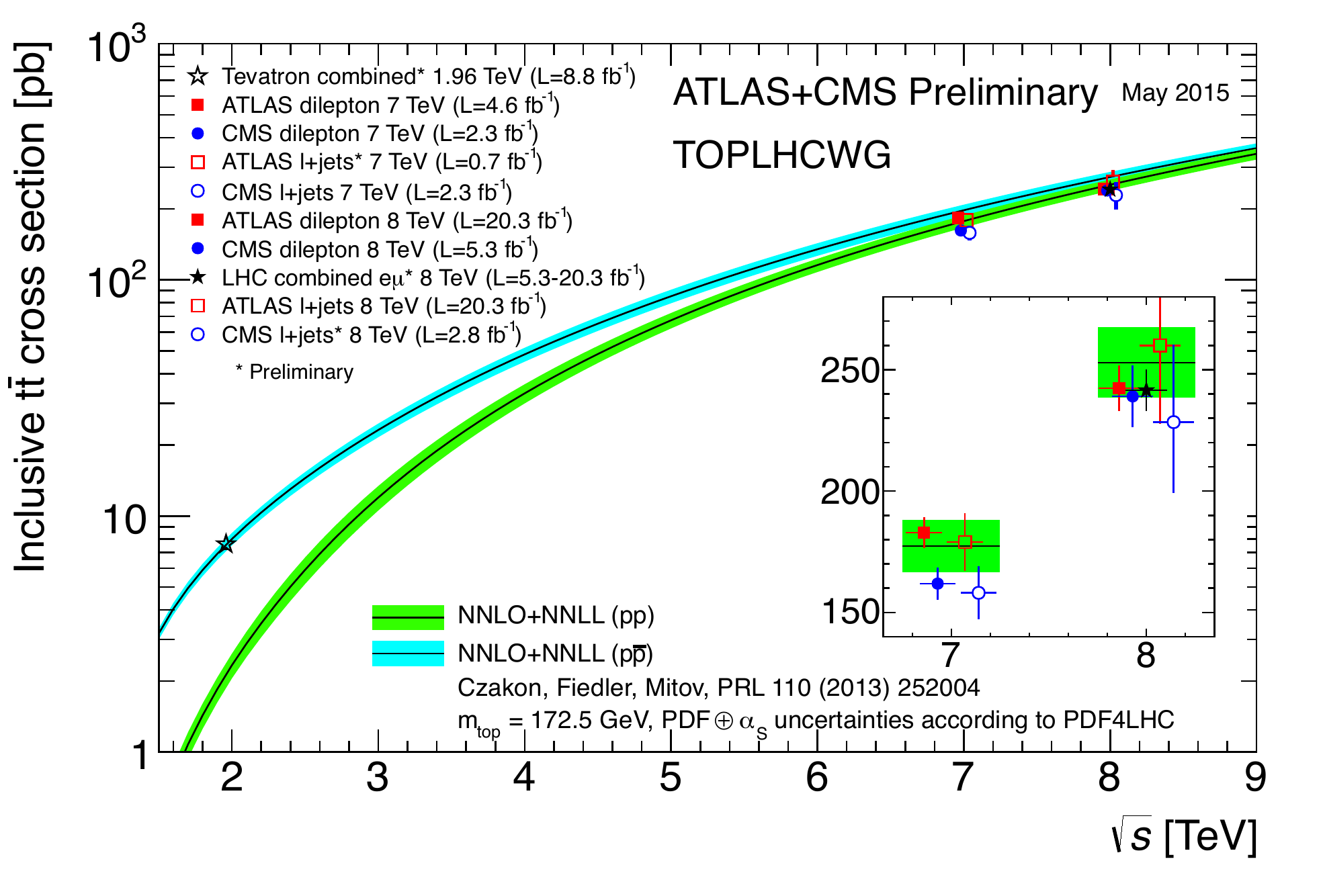}{0.88}{0}
\caption{Summary of \ttbar production cross-section measurements at the 
LHC~\cite{toplhcwg}
at 7 \TeV\ (top left) and 8 \TeV\ (top right), and 
most precise measurements in each
channel and experiment, together with the Tevatron and LHC combinations, as 
a function of the collision energy~\cite{toplhcwg} (bottom).  \label{tt_summary}}
\end{figure}

The best measurements of the inclusive \ttbar cross section are performed in the
dilepton $e\mu$ channel, as the background from multijet and vector-boson
production can be reduced to a negligible level.  Within the TOPLHCWG the best
8 \TeV\ measurements have been combined~\cite{combtt}, 
allowing for a reduction of systematic
uncertainties. The combined result is $\stt = 241.5 \pm 1.4 \stat
\pm 5.7 \syst \pm 6.2 \lumi \pb$, assuming a top-quark mass of $172.5 \GeV$.
All measurements are compatible with the theoretical predictions
(Fig.~\ref{tt_summary}).

\begin{figure}[!h]\centering
\subfig{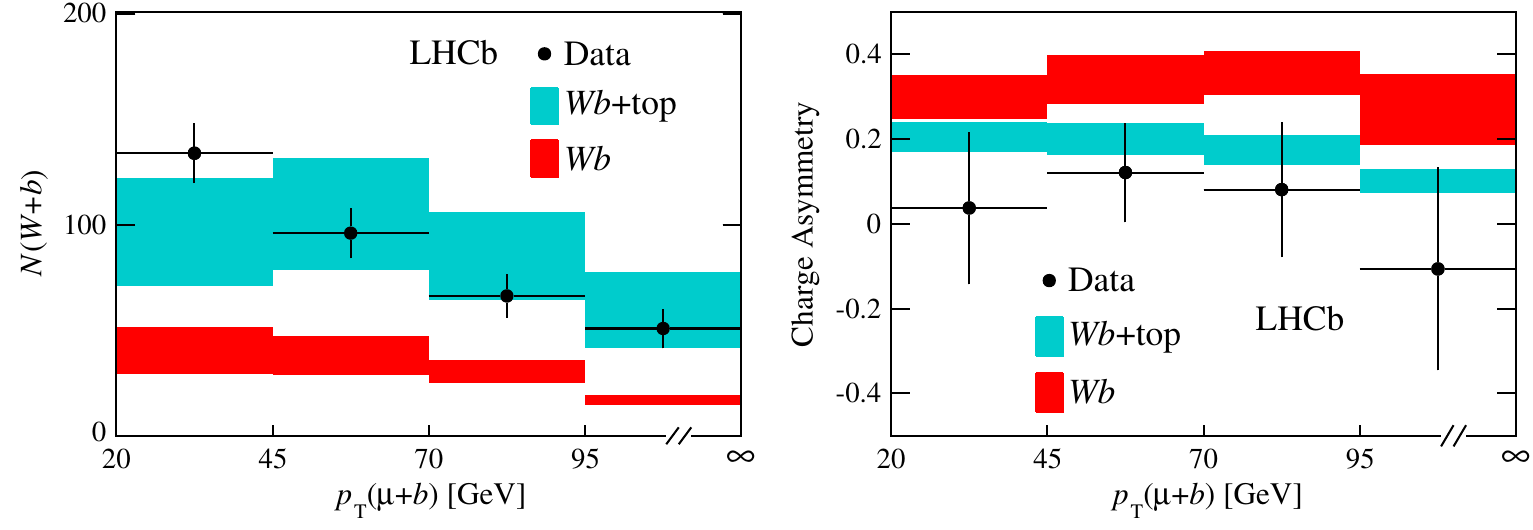}{0.99}{0}

\caption{Observation of top quarks in the forward region~\cite{lhcb}.  Results
for the $Wb$ yield (left) and charge asymmetry (right) as a function of the
transverse momentum of the $\mu b$-system, compared to 
data-constrained $Wb$ and NLO top predictions. \label{lhcb}}
\end{figure}

The LHCb experiment reported observation of top quarks in the forward
region~\cite{lhcb}. Albeit with large uncertainties this measurement is
interesting in its own, as it probes a complementary region of phase space,
possibly with enhanced sensitivity to new physics. The charge asymmetry 
and the number of events with a high-\pt muon and a $b$-tagged jet are
compared, as a function of \pt, to the hypothesis of $Wb$ production
(Fig.~\ref{lhcb}).  The result is obtained in a fiducial region, with $\pt(\mu)
> 25 \GeV$, $2.0 < \eta(\mu) < 4.5$, $\pt(b) > 50 \GeV$ and $2.2 < \eta(b) <
4.2$.  The likelihood fit in both variables shows that the background-only
hypothesis is excluded at the $5.4\,\sigma$ level.  Fiducial \ttbar cross
sections are extracted at 7 and 8 \TeV\ with total uncertainties of
$20$--$30\%$ and found to be consistent with SM expectations.

\begin{figure}[!h]\centering
\subfig{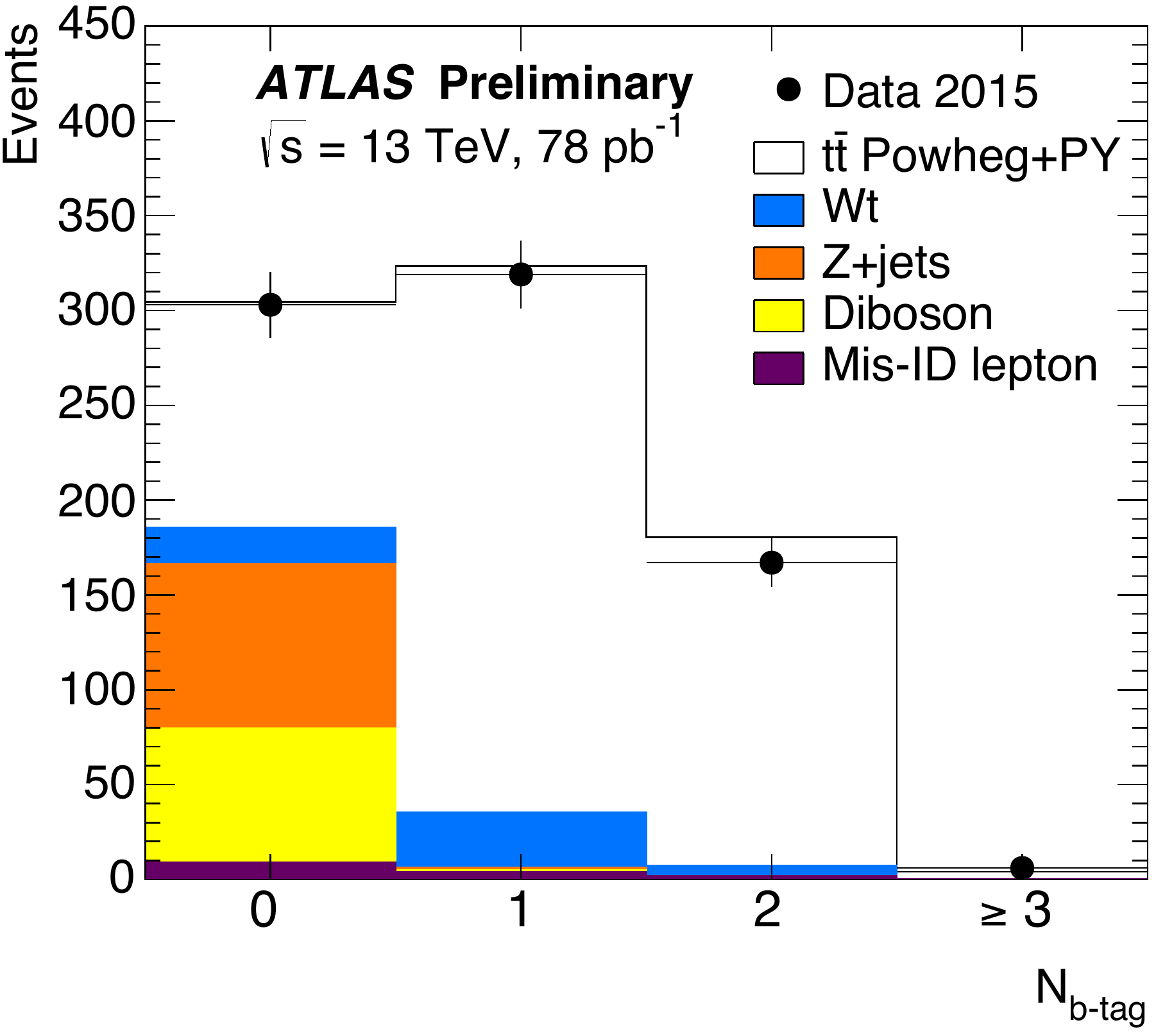}{0.407}{0}
\hspace{3ex}
\subfig{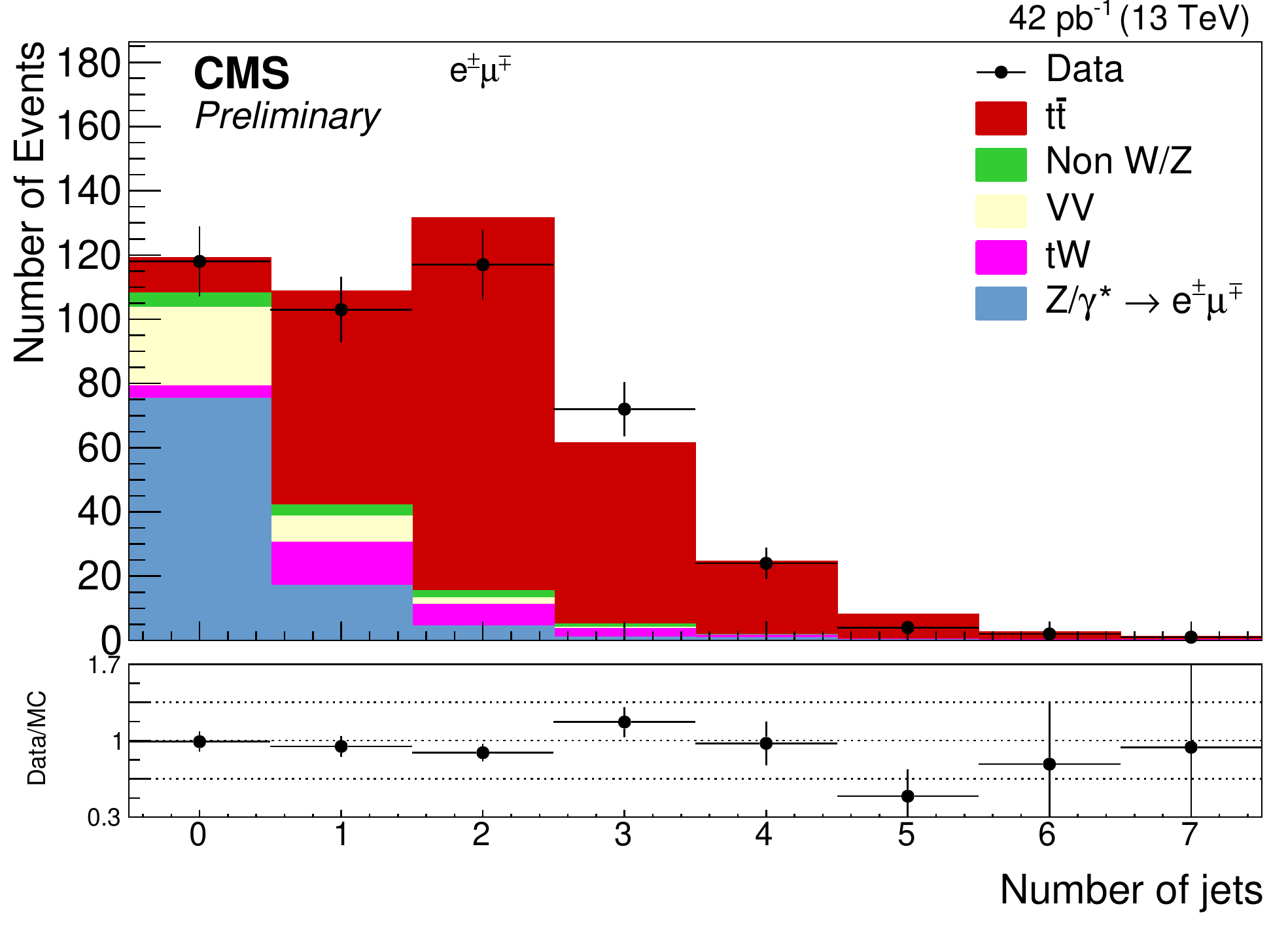}{0.483}{8}
\caption{Measurement of the \ttbar cross section at a centre-of-mass energy of 13 \TeV.
Number of $b$-tagged jets~\cite{a13} (left) and number of jets~\cite{c13} (right) 
for selected $e\mu$ events by ATLAS and CMS.\label{thirteen}}
\end{figure}

ATLAS and CMS started analysing proton--proton collision data at 13
\TeV, reporting first searches and measurements of Standard Model (SM) processes
\cite{beate, luca}.
Within few weeks of data-taking, events compatible with \ttbar dilepton, single-lepton
and single top quark production have been identified. With $\sim 80\;\ipb$ ATLAS
determines the \ttbar inclusive cross section closely following the Run-1 strategy
with $e\mu$ events~\cite{a13}. The number of events with one or two $b$-tagged jets are
used to simultaneously extract the cross section and the $b$-tagging efficiency.
The CMS analysis employs this same channel, with a smaller dataset of $\sim 40\;
\ipb$, but without using the information of $b$-tagging and is based on a simple
event counting technique~\cite{c13}. Besides the uncertainty on the integrated luminosity,
that affects the measurements at the level of $9$--$12\%$, the most important
systematic uncertainties are the modelling of the \ttbar hadronisation for ATLAS
($4.5\%$) and the uncertainty deriving from the lepton triggers for CMS
($5\%$). The result of these first two measurements is (Fig.~\ref{thirteen}):

\begin{eqnarray*}
\stt\mathrm{(ATLAS)} &=& 825 \pm 49 \stat \pm 60 \syst \pm 83 \lumi\; \mathrm{pb},\\
\stt\mathrm{(CMS)} &=& 772 \pm 60 \stat \pm 62 \syst \pm 93 \lumi\; \mathrm{pb}.\\
\end{eqnarray*}

\FloatBarrier
\section{Differential \ttbar cross sections}

With the large top-quark sample available at the LHC it is now possible to
study differential distributions. On one hand this allows for more detailed tests of
perturbative QCD, to constrain the parton distribution functions (PDF) and the
parameters of the Monte-Carlo simulation programs, on the other hand 
it allows for a better understanding of one of the major backgrounds in
Higgs physics, rare processes or search for beyond SM (BSM) effects. 
The strategy for differential measurements is to start with a
tight event selection to obtain a pure sample, enhanced with event
reconstruction techniques. After the estimated background is subtracted, the
effects of detector acceptance and resolution are removed by means of
unfolding. Distributions are then presented at parton or particle level, as a
function of kinematic quantities like \pt and $y$ of the \ttbar system or of
the top quarks, or of invariant masses.

\begin{figure}[!h]\centering
\subfig{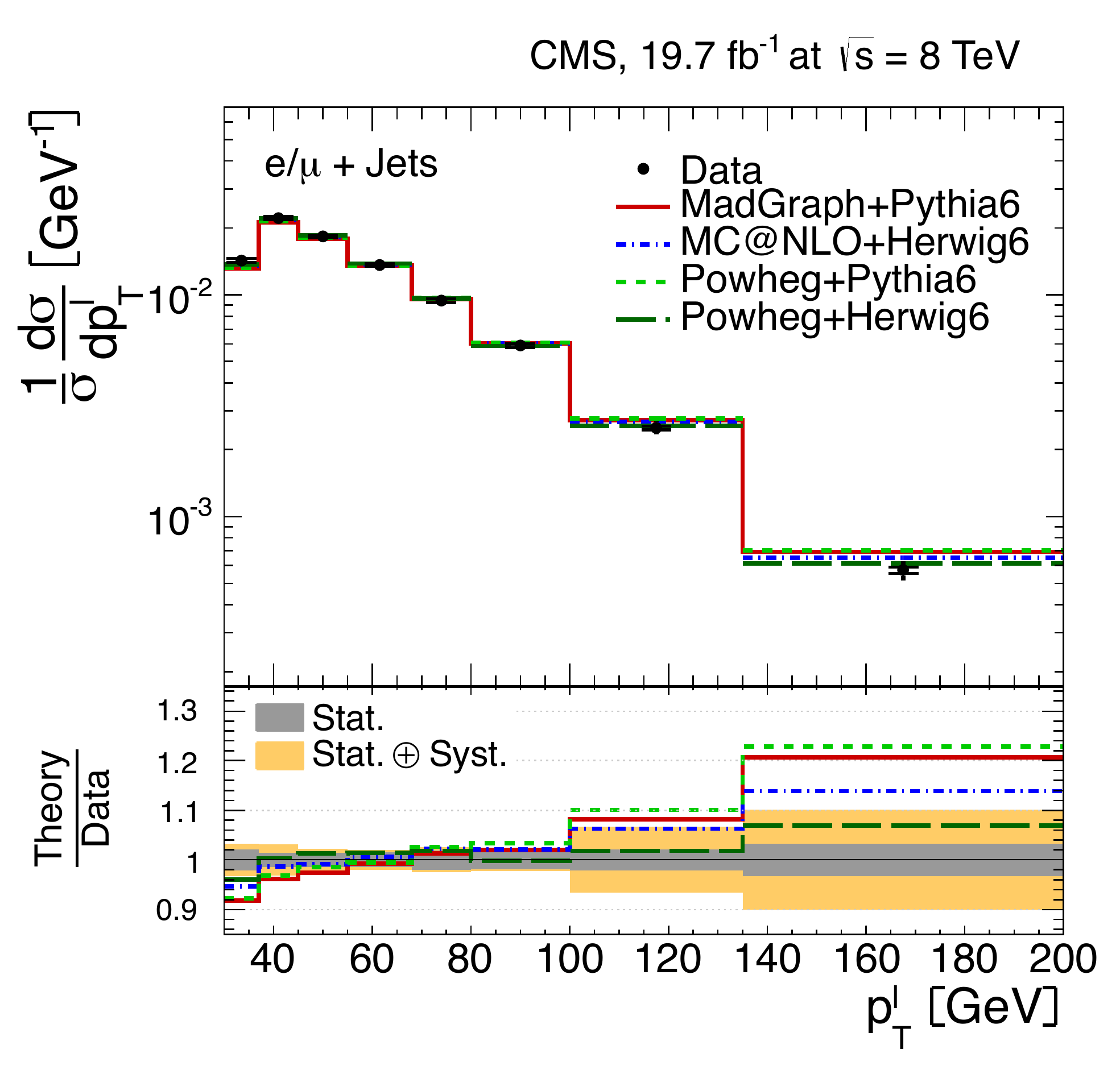}{0.43}{0}
\hspace{5ex}
\subfig{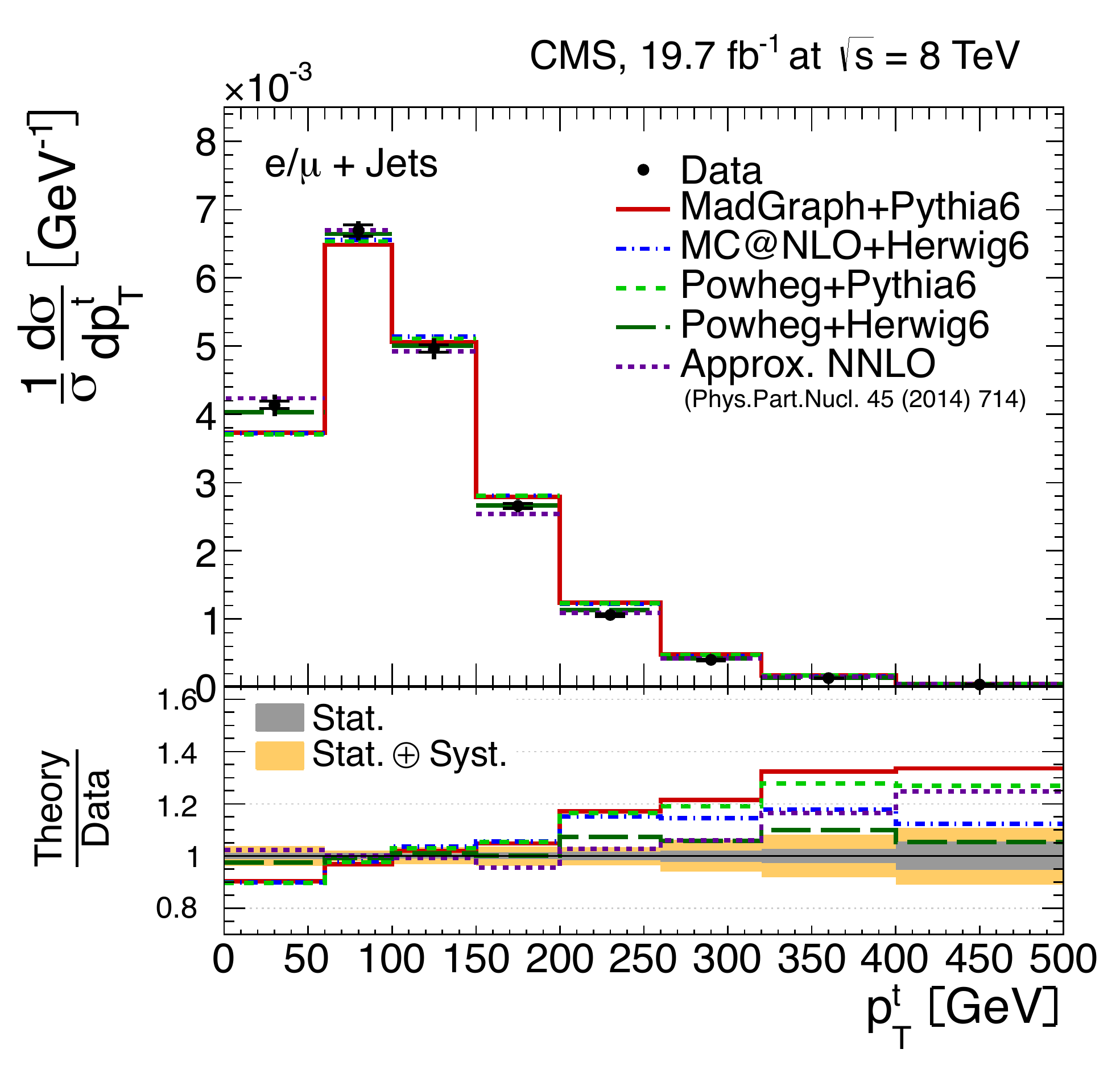}{0.43}{0}
\subfig{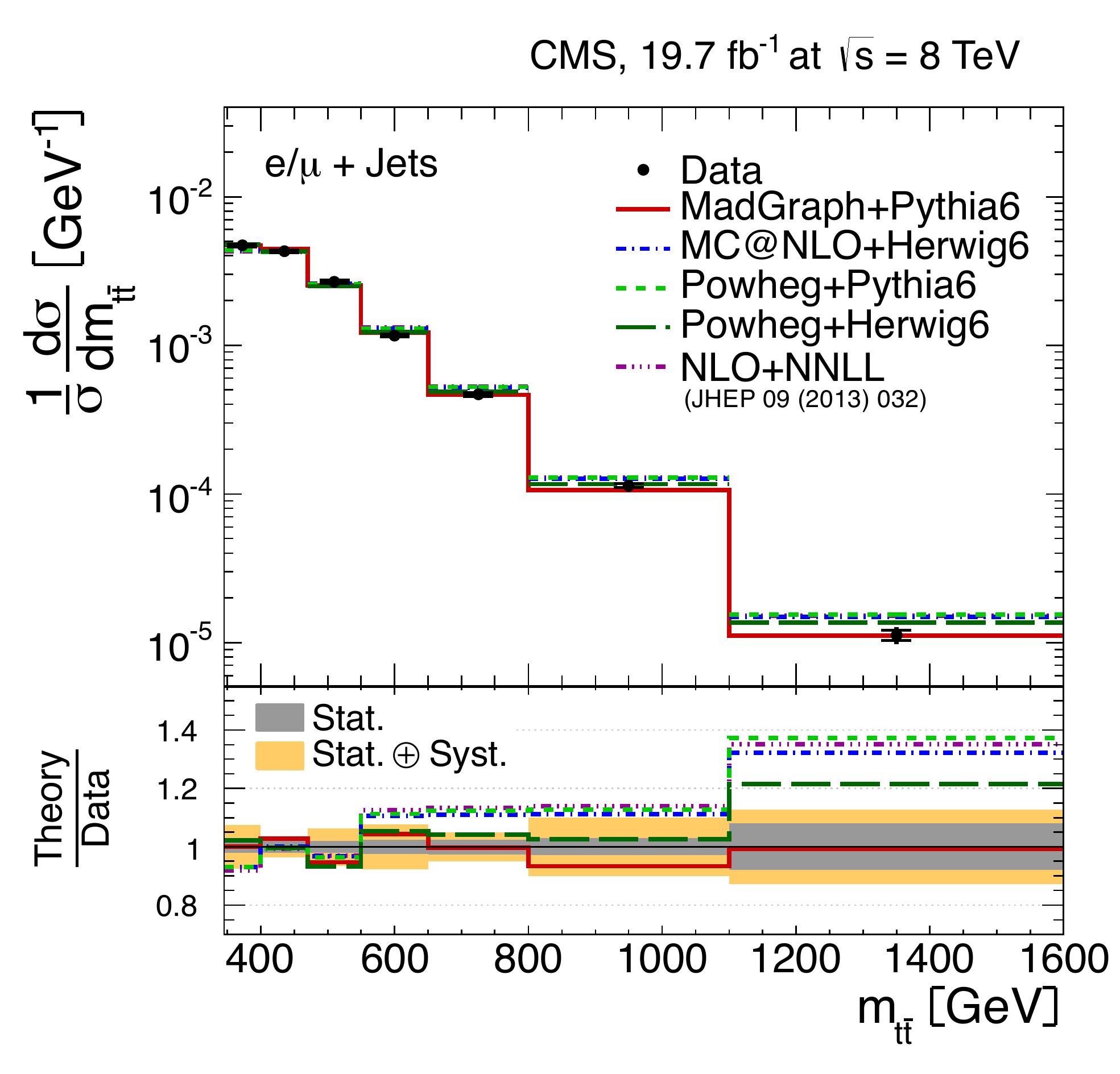}{0.43}{0}
\hspace{5ex}
\subfig{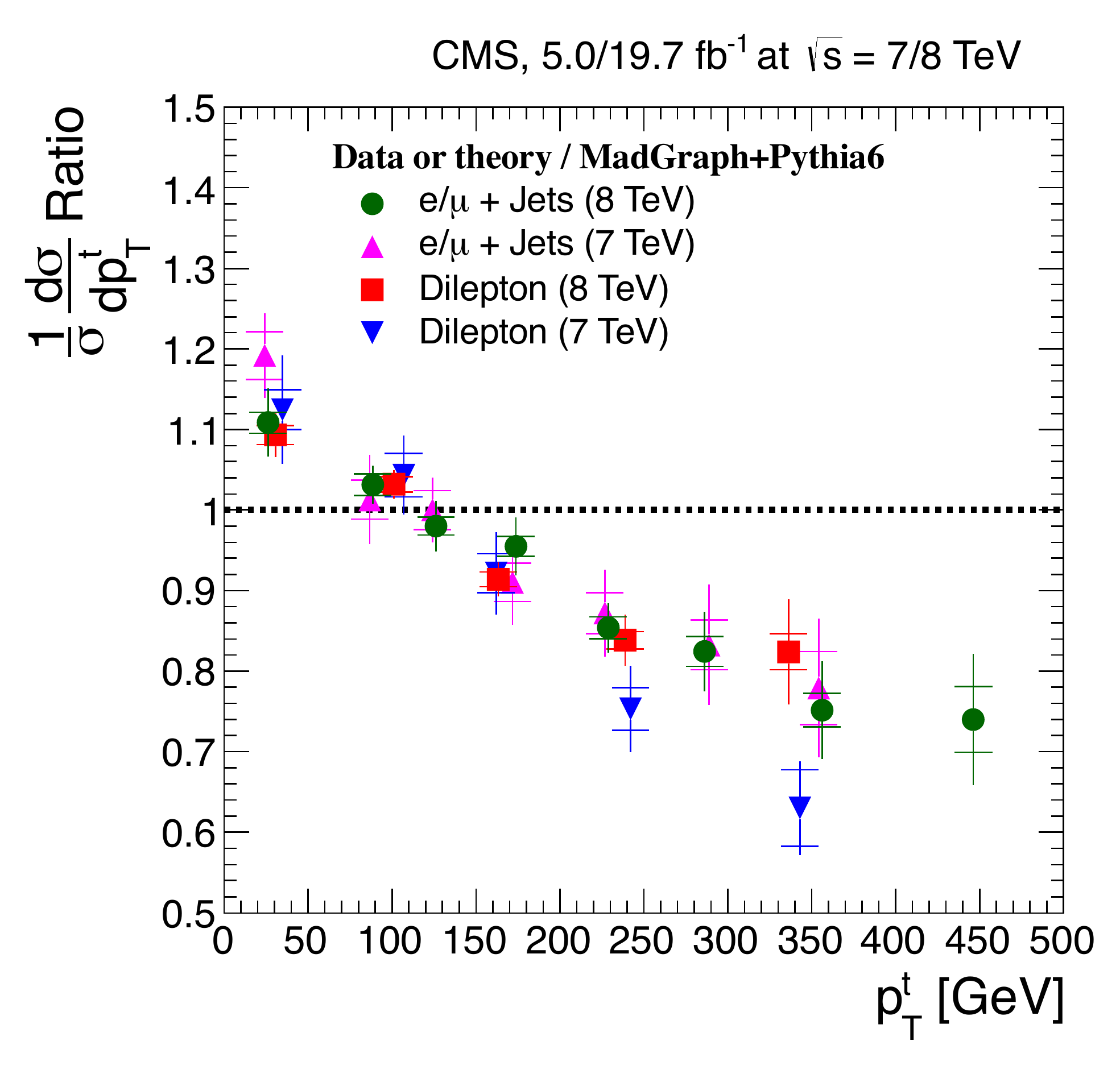}{0.43}{0}
\caption{Normalised differential \ttbar production cross section in the $\ell$+jets
channels compared to simulation~\cite{cdiff8}, 
as a function of the transverse momentum of the 
lepton (top left) and of the reconstructed top quark (top right), and the invariant mass
of the \ttbar system (bottom left). Summary of all measurements as a function of
top \pt~\cite{cdiff8} (bottom right). \label{differential}}
\end{figure}

A comprehensive set of top-quark differential measurements at $8 \TeV$ has
been performed by CMS using final states with leptons~\cite{cdiff8}. The
comparison to different matrix-element and parton-shower programs is provided
for the top-quark decay products, as well as for reconstructed objects at
parton level. This analysis confirms the trend observed at $7 \TeV$ that
the NLO simulation does not fully describe the \pt distribution of the top
quark (Fig.~\ref{differential}). 

\begin{figure}[!h]\centering
\subfig{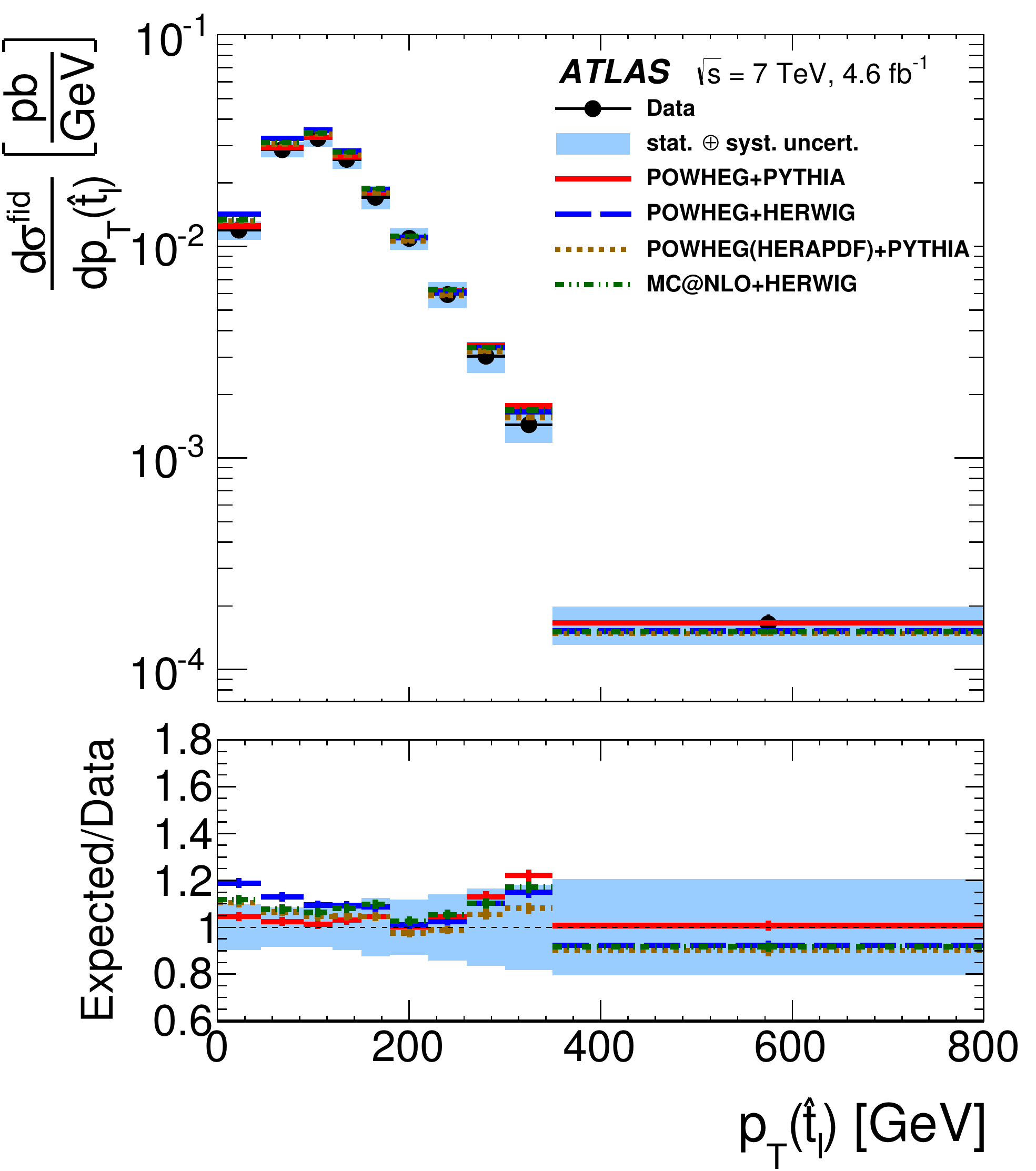}{0.325}{0}
\subfig{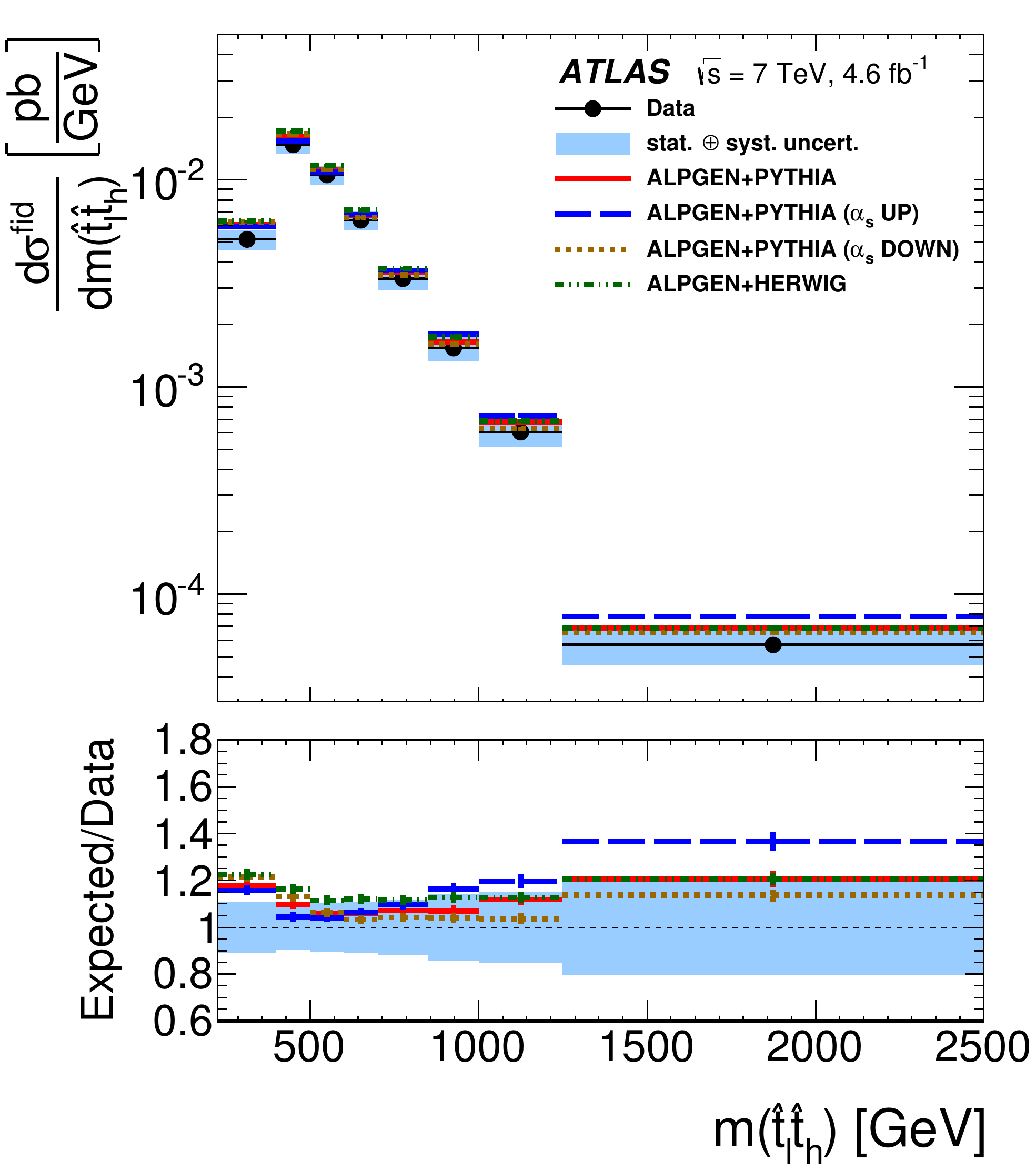}{0.325}{0}
\subfig{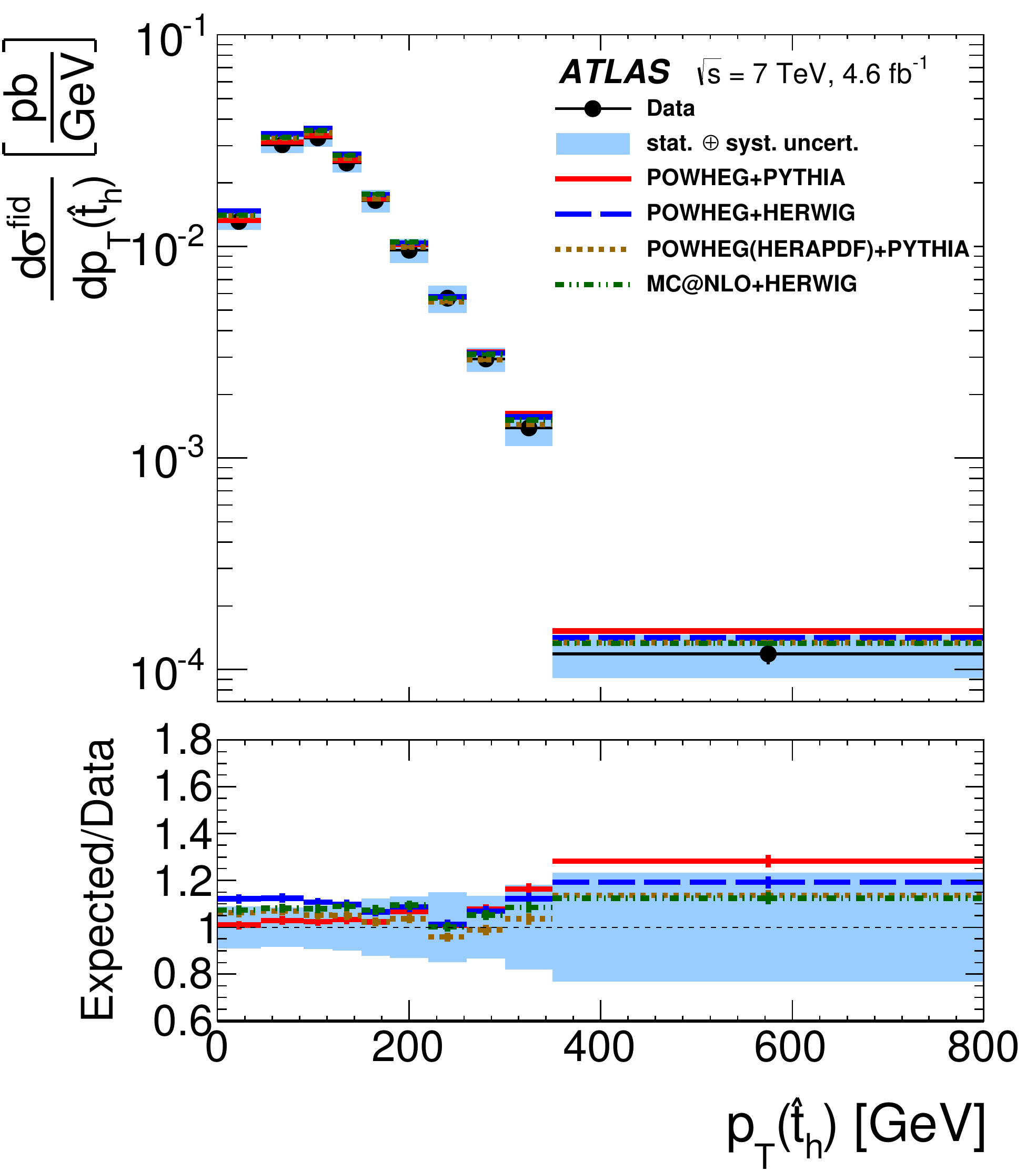}{0.325}{0}
\caption{Differential \ttbar cross section~\cite{pseudo} as a function of the 
hadronic (left)
and leptonic (centre) pseudo-top-quark \pt, and of the invariant mass of the two
pseudo-top quarks (right). Model predictions from several generators are 
superimposed. \label{pseudo}}
\end{figure}

Differential measurements presented at parton level are model dependent. In an
effort to reduce theoretical uncertainties ATLAS made use of the concept of the 
so-called pseudo-top quark, a proxy object directly constructed from
detector-level observables: charged leptons, jets and missing transverse
momentum. The analysis of $7 \TeV$ data~\cite{pseudo} reveals a good description
of the data in general, with some discrepancies at low $m_{\ttbar}$ values.
Within uncertainties the top-\pt distribution is sufficiently well described by 
\textsc{Powheg+Pythia} (Fig.~\ref{pseudo}).

The high top-quark \pt region has been the target of further detailed studies, since
a possible discrepancy could hint to physics effects beyond the Standard Model
and studies allow for an improved understanding of the proton PDF.
When top quarks are produced with large Lorentz boost their decay products tend 
to be more collimated and might escape standard reconstruction techniques, which
exploit isolation. In particular, the hadronically decaying top quark is 
reconstructed as a single large-radius ($R$) jet. Jet substructure techniques are
employed to identify such large-$R$ jets and tested for compatibility 
with top-quark decays.

\begin{figure}[!h]\centering
\subfig{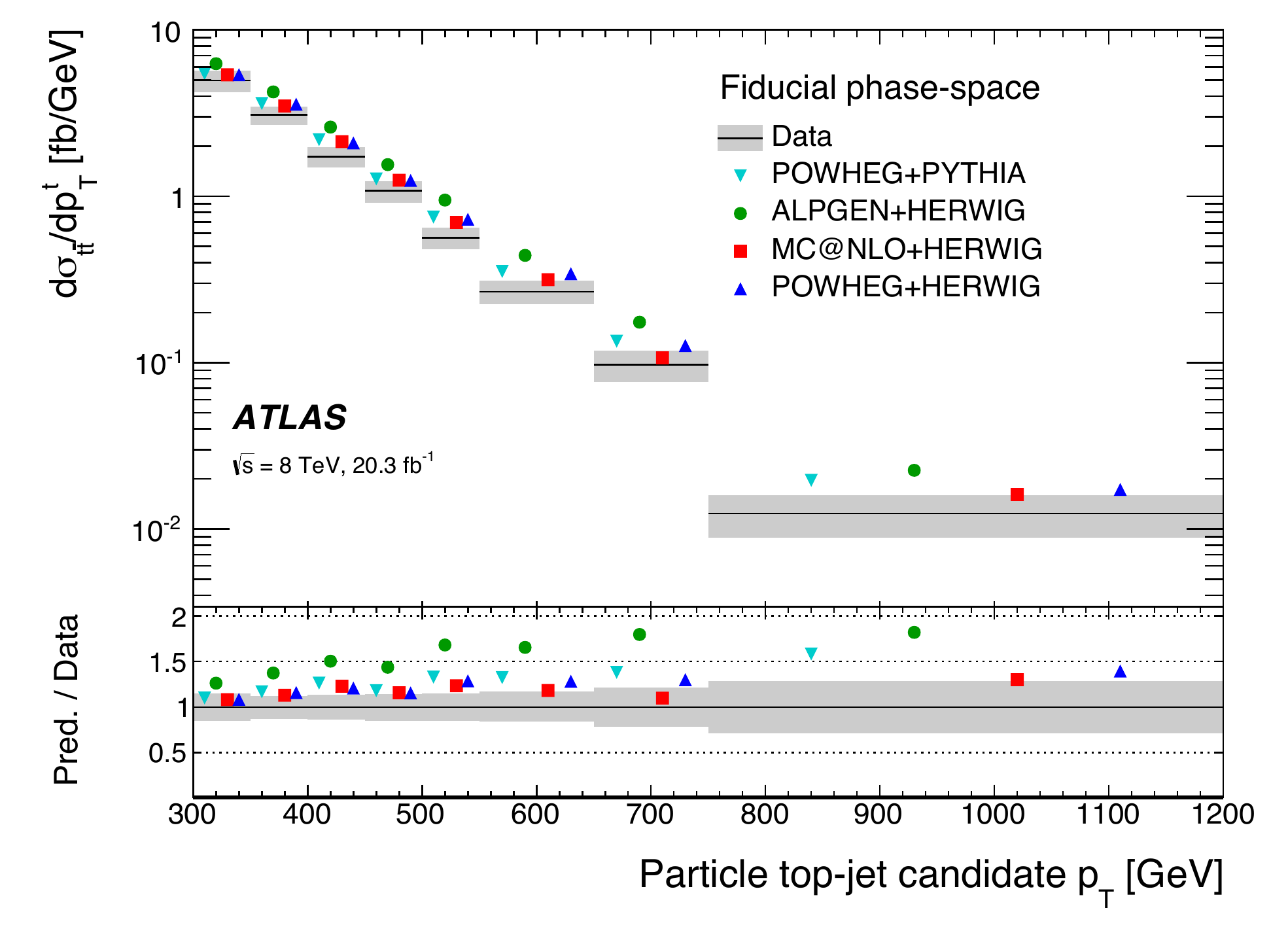}{0.44}{0}
\hspace{4ex}
\subfig{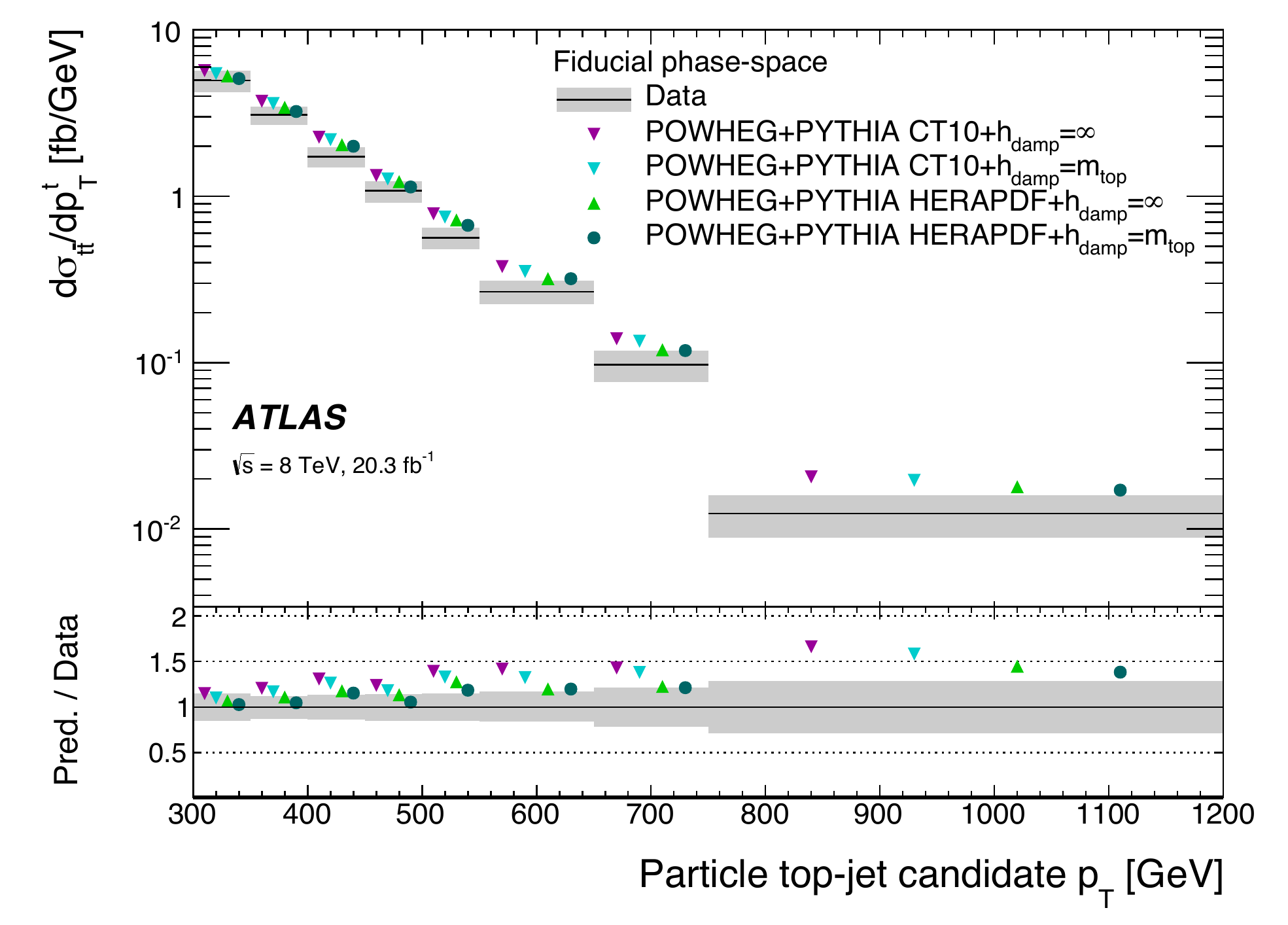}{0.44}{0}
\subfig{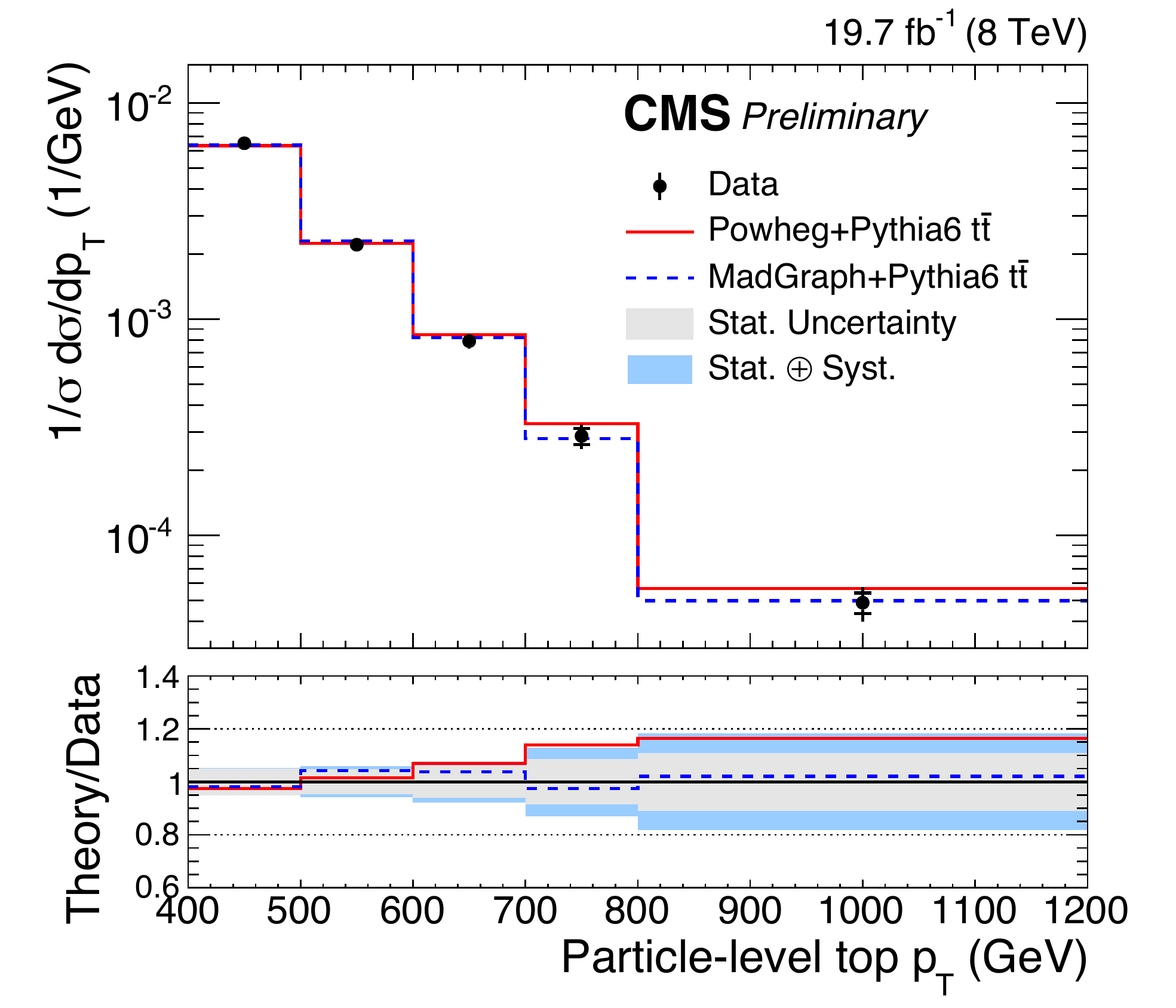}{0.4}{0}
\hspace{6ex}
\subfig{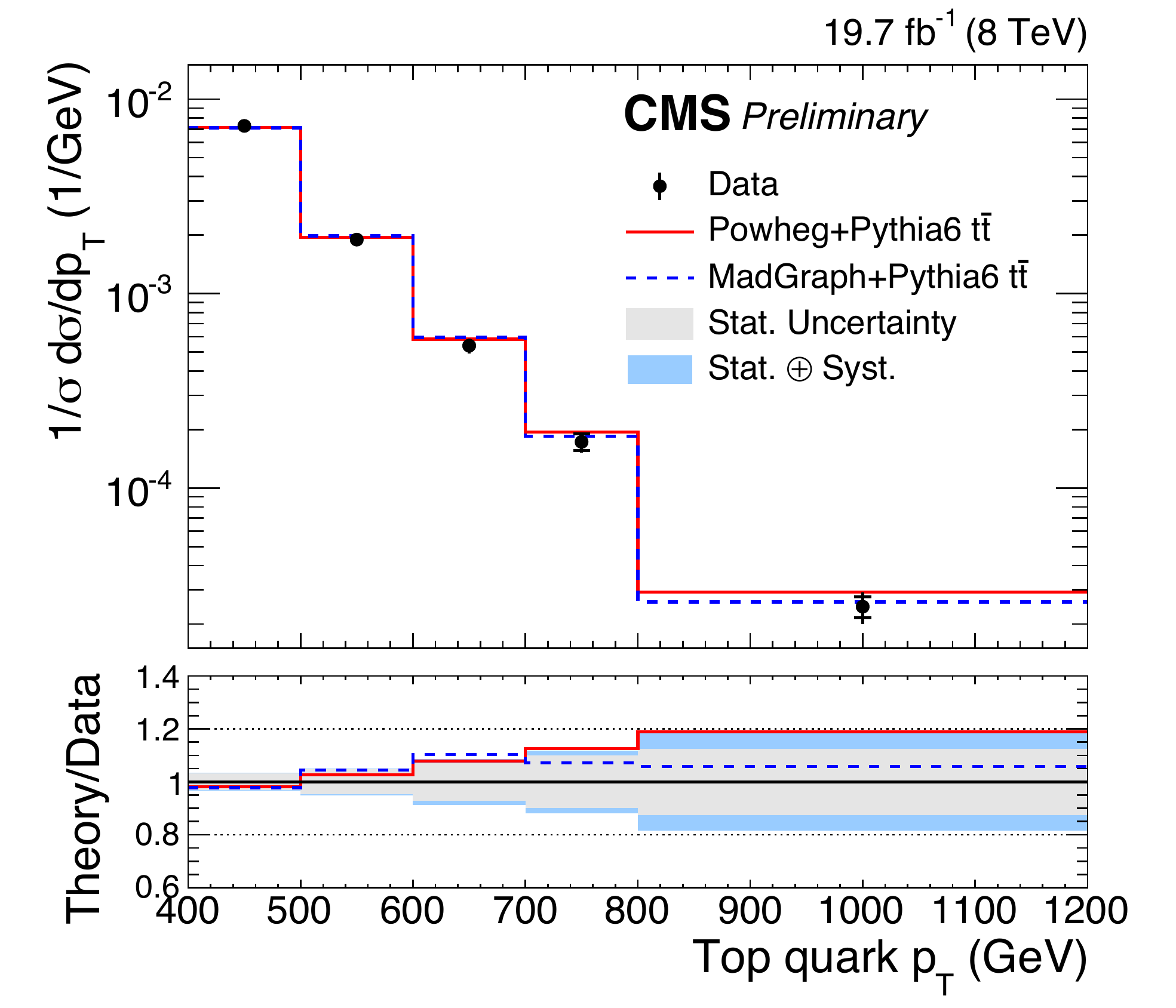}{0.4}{0}
\caption{Fiducial particle-level differential cross section as a function of
the hadronic top-jet candidate \pt reconstructed with boosted techniques~\cite{aboo}, 
compared to LO and NLO generators interfaced to different parton-shower programs 
(top left) and to \textsc{Powheg+Pythia} with
varying shower parameter $h_\mathrm{damp}$ or PDF (top right). Normalised differential
cross section~\cite{cboo} in bins of the particle-level top-jet \pt (bottom left) and the
parton-level top-quark \pt (bottom right).\label{boosted}}
\end{figure}

Measurements are performed with the full $8 \TeV$ dataset.
ATLAS selects jets with $R=1$ and measures differential \ttbar cross sections
as a function of the top-quark \pt both, at particle level in a fiducial
region closely following the event selection and at parton level~\cite{aboo}.
The measurements have a threshold of $300 \GeV$ and extend beyond 
$1 \TeV$. Experimental uncertainties are of the order of $10$--$30\%$ and
are dominated by the jet energy scale uncertainty of large-$R$ jets.
Discrepancies between data and simulations observed at low-$\pt$ are in general
confirmed in the boosted regime, although not statistically significant.
Different PDF and parton-shower parameter settings can improve
the agreement between data and simulation considerably (Fig.~\ref{boosted}).
CMS performs a similar analysis with $R=0.8$ jets to extract normalised
differential distributions starting from $400 \GeV$~\cite{cboo}.

\section{Single-top quark production}

Besides the dominating pair-production of top quarks through the strong
interaction, top quarks can also be produced singly via the weak vertex
$Wtb$. There are three possible single top-quark production modes at
leading order in perturbation theory: an exchange of a virtual $W$ boson either
in the $t$-channel or in the $s$-channel, or the associated production of a
top-quark and a $W$ boson.

\begin{figure}[!h]\centering
\subfig{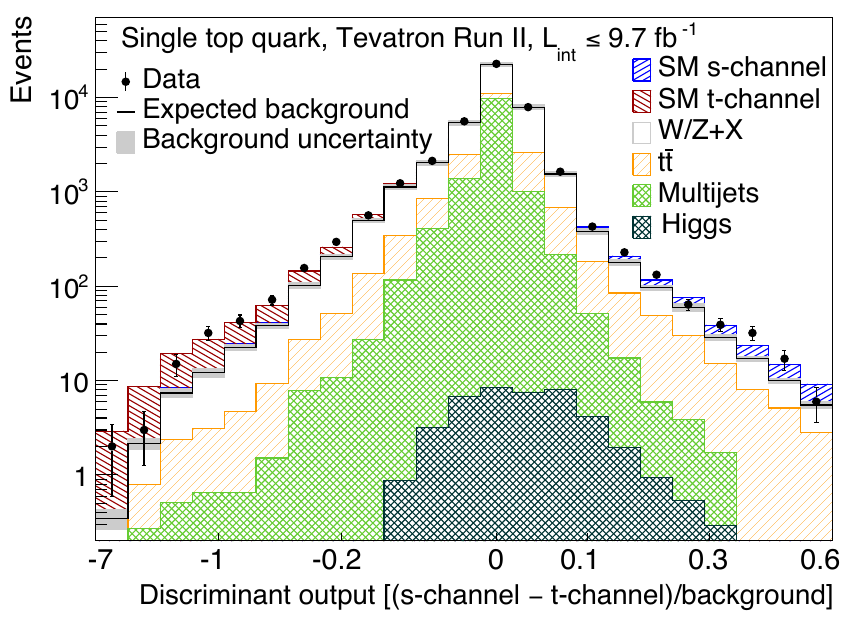}{0.48}{8}
\hspace{5ex}
\subfig{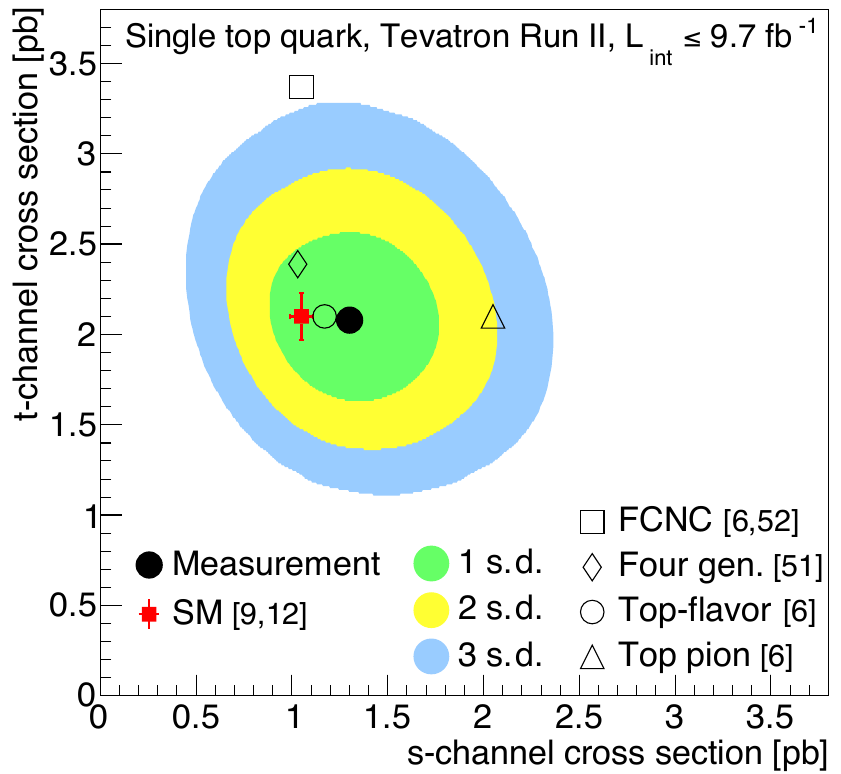}{0.4}{0}
\caption{Distribution of the mean discriminants for bins with similar ratios of 
($s$-channel $-$ $t$-channel) signals divided by background yields~\cite{tevst} 
(left) and
two-dimensional posterior probability as a function of $\sigma_t$ and 
$\sigma_s$ with probability contours for the combination of the CDF and D0 
analysis channels compared with the theoretical prediction of the SM and
several BSM predictions~\cite{tevst} (right). \label{tev-st}}
\end{figure}

Much progress has been reported in the studies of single-top quark production
in the three channels at Tevatron and LHC. In a final combination of the
Tevatron datasets~\cite{tevst} the inclusive single-top $t$-channel
cross section is measured with a relative uncertainty of $13\%$. At the same
time the $s$- and $t$-channel production cross sections are simultaneously
extracted in a two-dimensional measurement.  A likelihood fit is performed to
the binned distribution of a discriminant, optimised to separate signal
events from large background contributions and to separate $s$-channel and
$t$-channel events. The two channels are sensitive to different new
physics effects, as can be seen in Fig.~\ref{tev-st} and it is thus important
to measure them separately.  The $Wt$ channel remains unaccessible at the
Tevatron.  

\begin{figure}[!h]\centering
\subfig{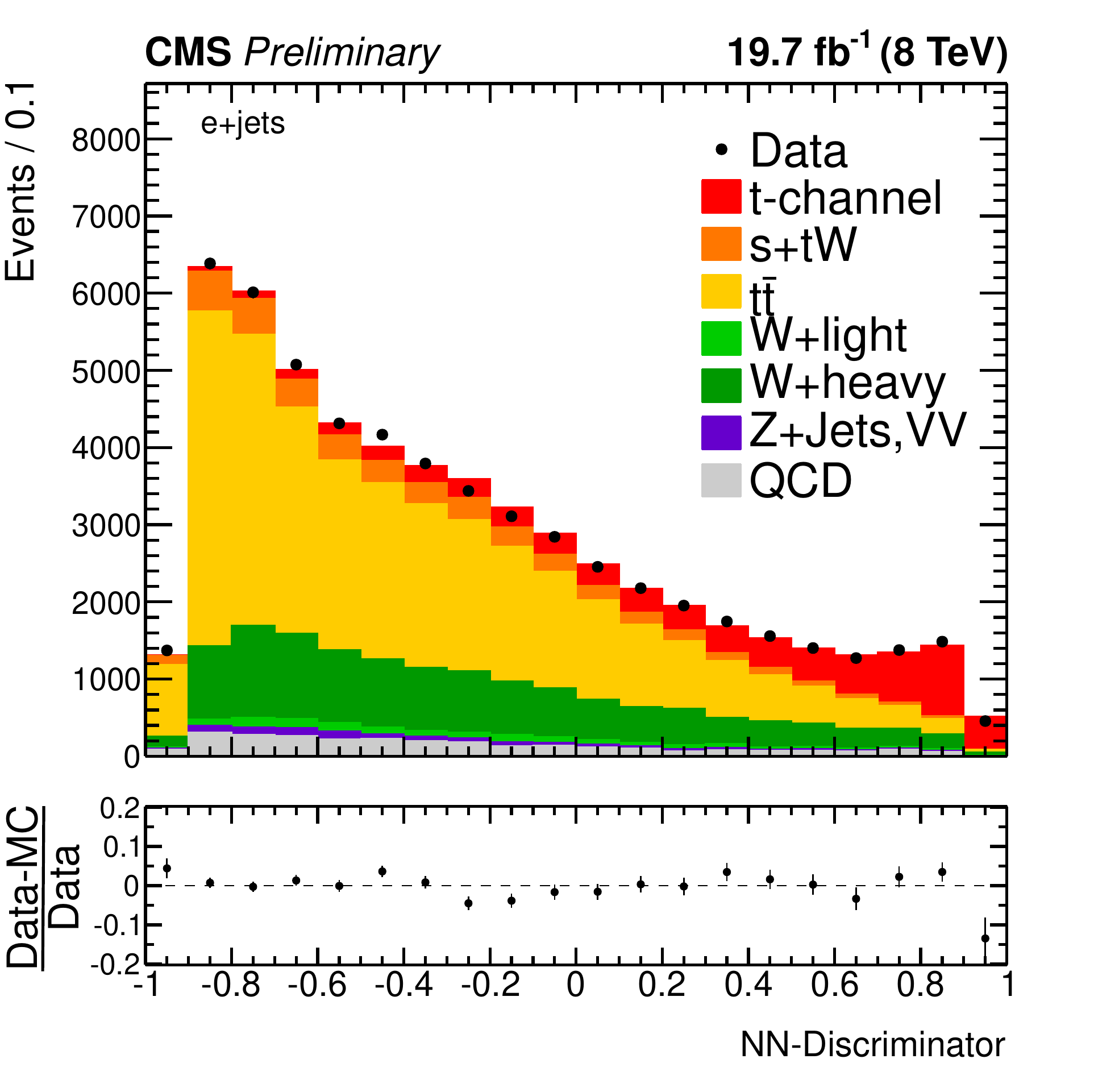}{0.325}{0}
\subfig{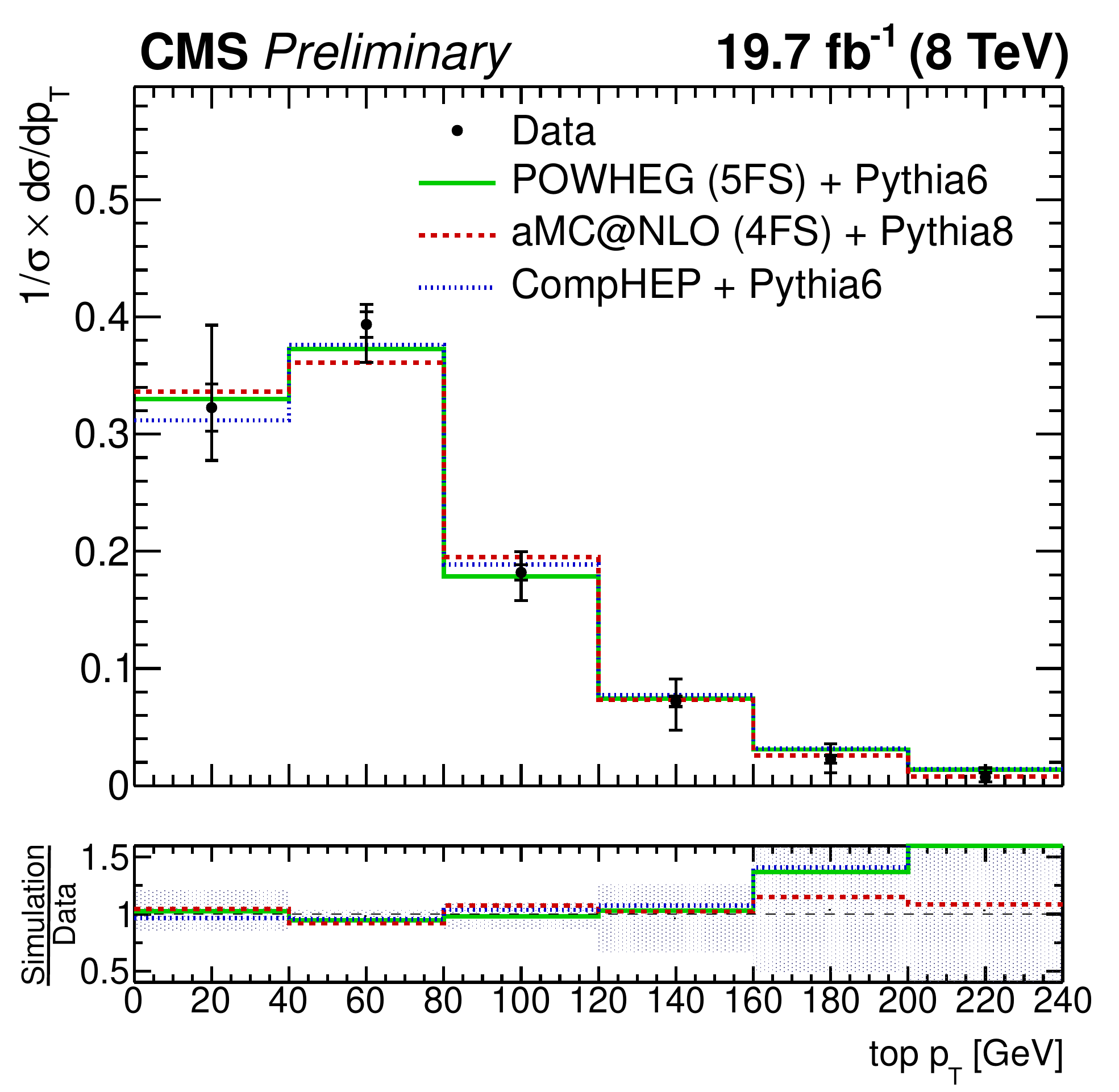}{0.325}{0}
\subfig{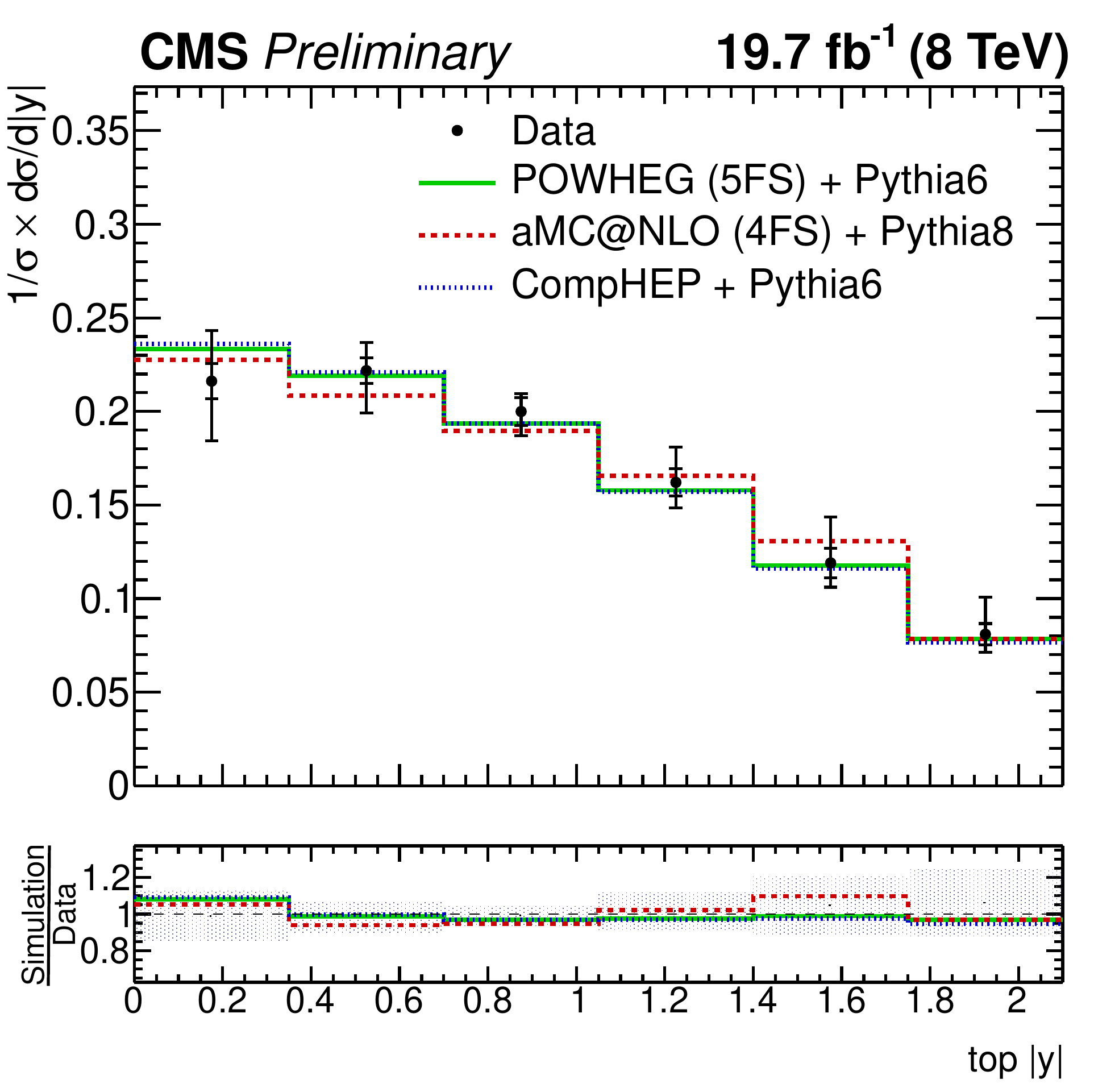}{0.325}{0}
\caption{Measurement of the single top $t$-channel cross section at 
8 \TeV~\cite{ctchan}:
neural-network output distribution in the $e$+jets channel, requiring
two jets and one $b$-tag (left), and unfolded spectra in the 
$\ell$+jets channel compared with predictions as a function of top quark \pt (centre)
and rapidity $|y|$ (right).  \label{t-channel8}}
\end{figure}

The large LHC dataset not only allows for precise measurements of the
inclusive single top $t$-channel cross section at 7 and 8 \TeV, but also for
the measurement of properties and differential distributions. The most recent
measurement is performed by the CMS Collaboration, employing a neural network
(NN) discriminator with variables like the pseudorapidity of the light quark
jet or the invariant mass of the reconstructed top-quark candidate $m_{\ell\nu
b}$~\cite{ctchan}.  After a stringent requirement on the NN discriminator
output, a good ratio $S/\sqrt{S+B}$ is achieved and normalised
background-subtracted and unfolded distributions as a function of the top-quark
\pt and $|y|$ are shown to be in good agreement with NLO generators
(Fig.~\ref{t-channel8}), using four- or five-flavour schemes (a\textsc{MC@NLO}
and \textsc{Powheg}), as well as \pt-matched samples, that simulate $2\to2$ and
$2\to3$ processes (\textsc{CompHEP}).

\begin{figure}[!h]\centering
\subfig{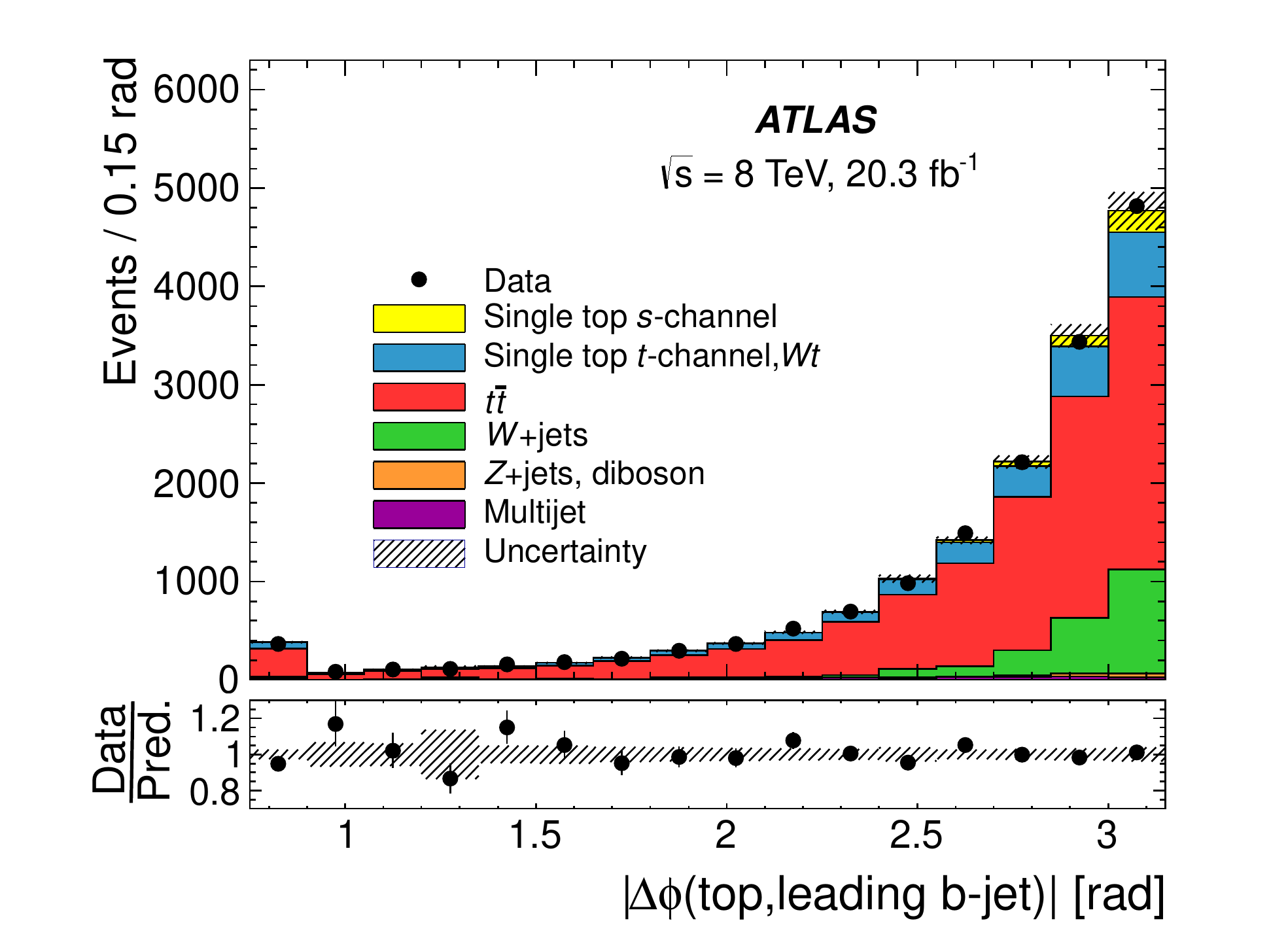}{0.45}{0}
\subfig{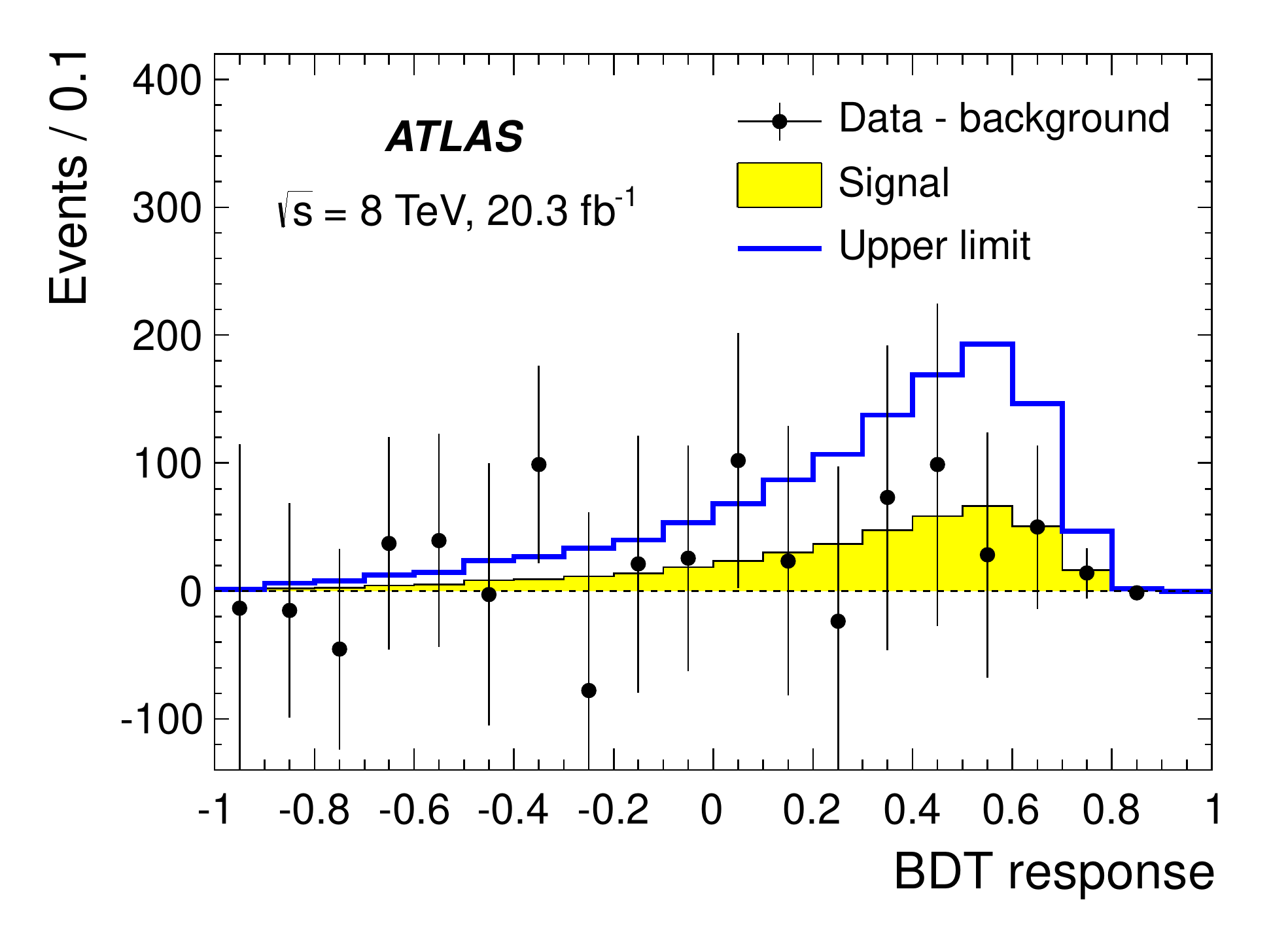}{0.45}{0}
\caption{Measurement of the single top $s$-channel cross section at 
8 \TeV~\cite{aschan}:
observed and predicted distributions in the signal region for the azimuthal angle 
between the leading $b$-jet and the top-quark candidate reconstructed with 
the sub-leading $b$-jet (left), and BDT response after background subtraction (right).
\label{s-channel}}
\end{figure}

At the LHC a measurement of the production in the $s$-channel is much more 
challenging, as it is suppressed by the requirement of a quark-antiquark pair
in the colliding protons with sufficiently large $x$. Thus, while at Tevatron the 
expected ratio of $t$-to-$s$ production is 2, at LHC at $8 \TeV$ this ratio is 15.
Consequently the signal-to-background ratio in this channel is quite prohibitive
at LHC. With the full dataset ATLAS performed a search for production in this
channel~\cite{aschan}. After event selection with one lepton and exactly two
jets, both of which are $b$-tagged, the signal-to-background ratio is $3\%$.
A boosted decision tree (BDT) classifier is trained with kinematic and
topological variables.
The most discriminating variables are the differences in azimuthal angle 
$|\Delta \phi|$ between the $b$-jet and the top-quark candidate. The BDT output
distribution is used to extract the signal contribution using a maximum
likelihood fit (Fig.~\ref{s-channel}). 
The observed significance of the measurement is found to be $1.3\,\sigma$
and the cross section is measured to be $\sigma_s = 5.0 \pm 4.3$ pb.

\begin{figure}[!h]\centering
\subfig{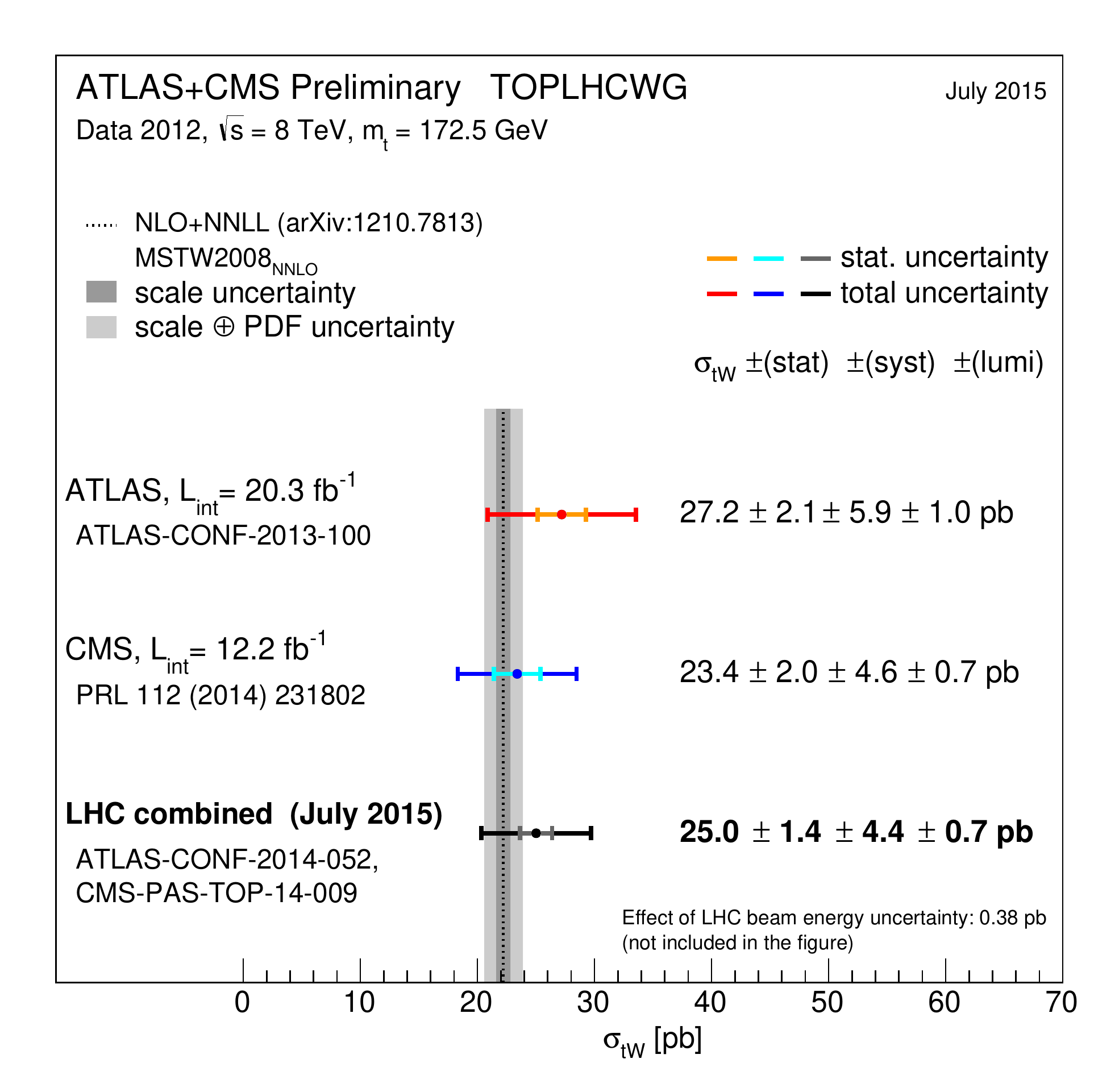}{0.51}{0}
\subfig{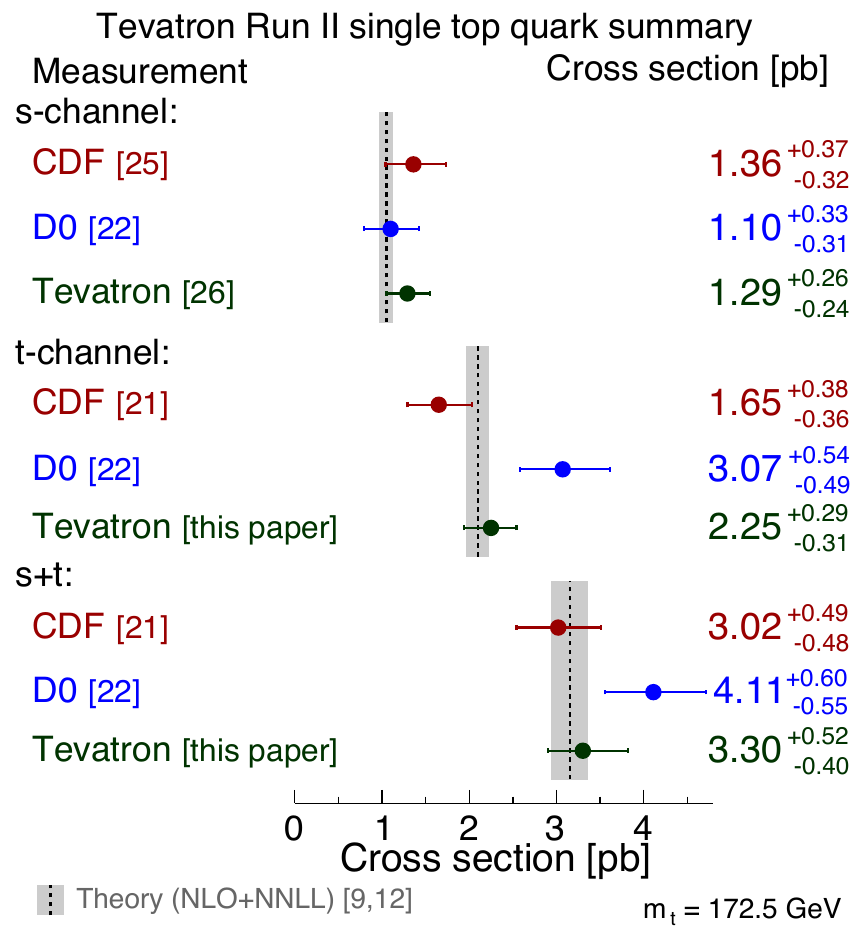}{0.47}{0}
\subfig{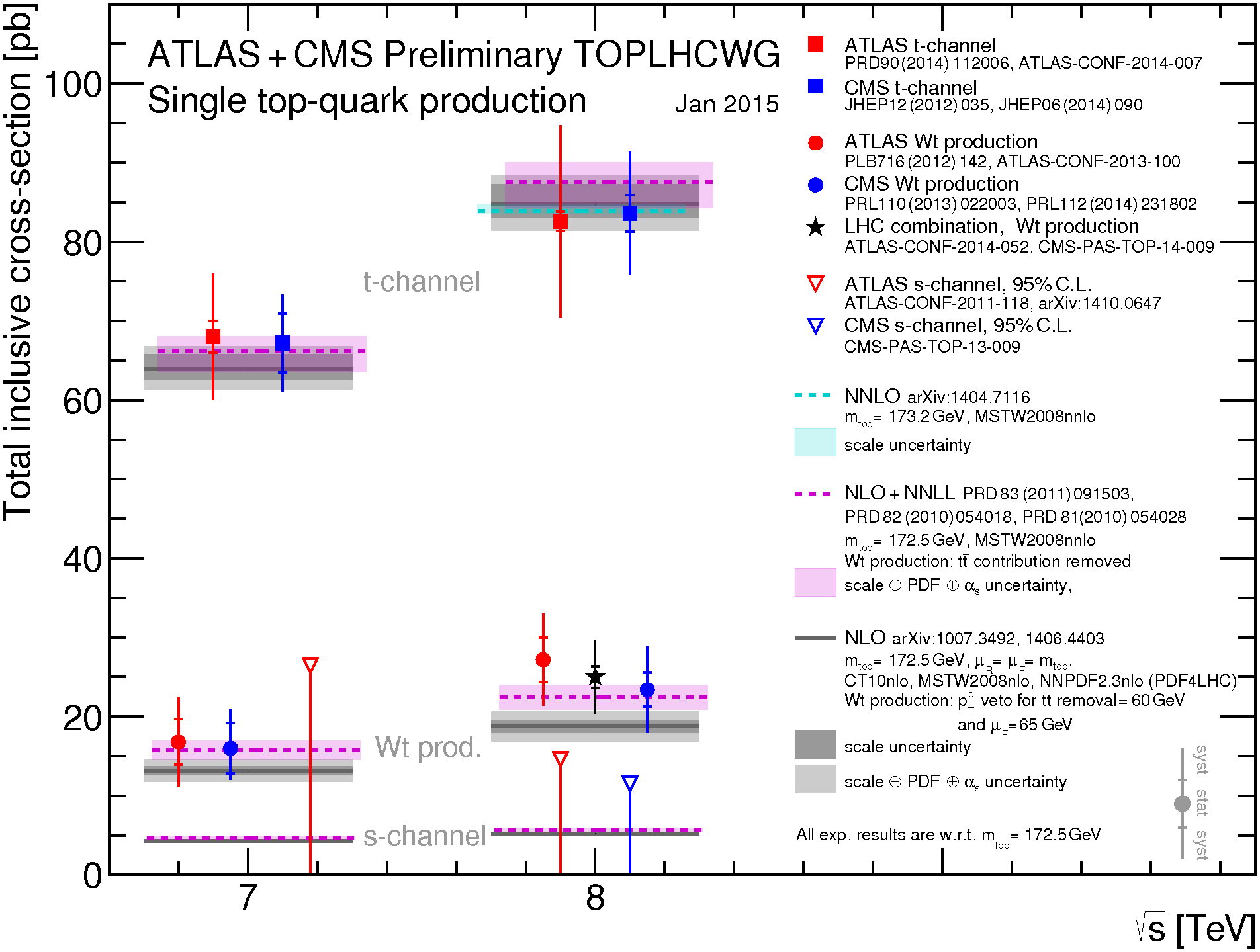}{0.75}{0}
\caption{Single top summary plots including the results of combinations and 
compared to predictions: TOPLHCWG $8 \TeV$ $Wt$ channel
combination~\cite{combWt} (top left), Tevatron Run II~\cite{tevst} (top right) and 
LHC summary plot~\cite{toplhcwg} (bottom).  \label{st-combi}}
\end{figure}

Measurements of single top production in the $Wt$ channel at the LHC have 
been combined within
the \mbox{TOPLHCWG~\cite{combWt}}. In both cases the dilepton signature with one
or two jets is exploited as it offers a good signal-to-background ratio. 
The final discrimination is obtained with BDTs. The CMS measurement uses all
dilepton channels and a partial dataset, while the ATLAS measurement employs
only the $e\mu$ channel and the full dataset. The total uncertainty of 
$23\%$ in each of the input measurements is reduced to $19\%$ in the combination, with
small statistical uncertainties (Fig.~\ref{st-combi}). 
The uncertainty due to the matching of 
matrix element and parton shower and the choice of scales dominate the
systematic uncertainty.

Each measurement of single top production is in good agreement with the 
Standard Model expectation (Fig.~\ref{st-combi}) and can be separately 
interpreted in terms of a determination of the CKM matrix element \Vtb. 
The Tevatron combination
extracts a value of $1.02^{+0.06}_{-0.05}$, while at the LHC the measurements
with the smallest uncertainties are derived from the CMS 7 and 8 \TeV\
combination for the $t$-channel ($1.00 \pm 0.04$) and the LHC combination for
the $Wt$ channel ($1.06 \pm 0.11$). 

\section{Associated production}

Several measurements of top-quark production in association
with further particles have been performed. 
Recent observation, evidence or search for production
of top quark pairs in association with photons, electroweak bosons, as well as
heavy quarks have been reported and are discussed in the following. The
production of \ttH is presented elsewhere~\cite{sinead}.

\begin{figure}[!h]\centering
\subfig{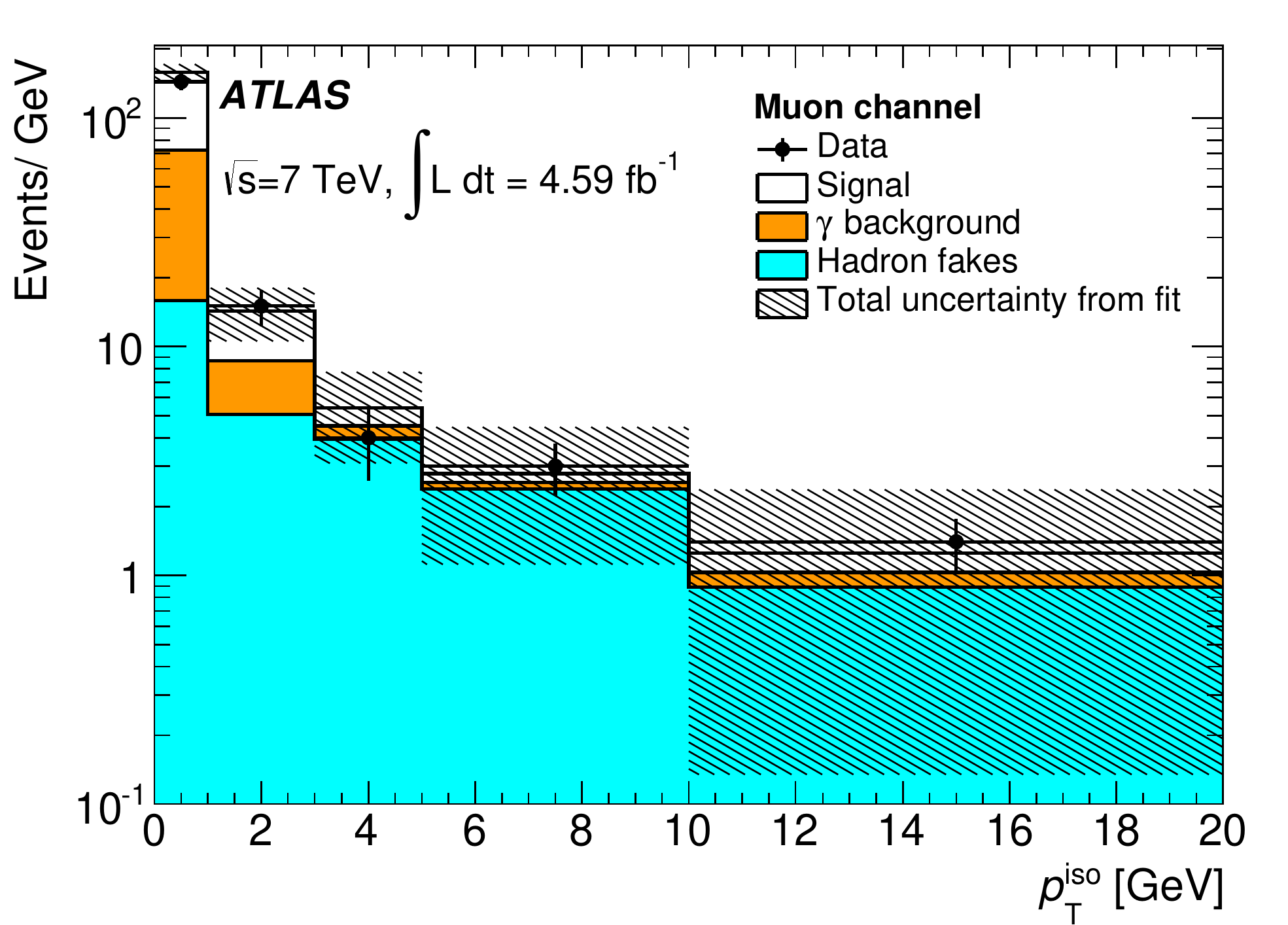}{0.46}{7}
\hspace{1ex}
\subfig{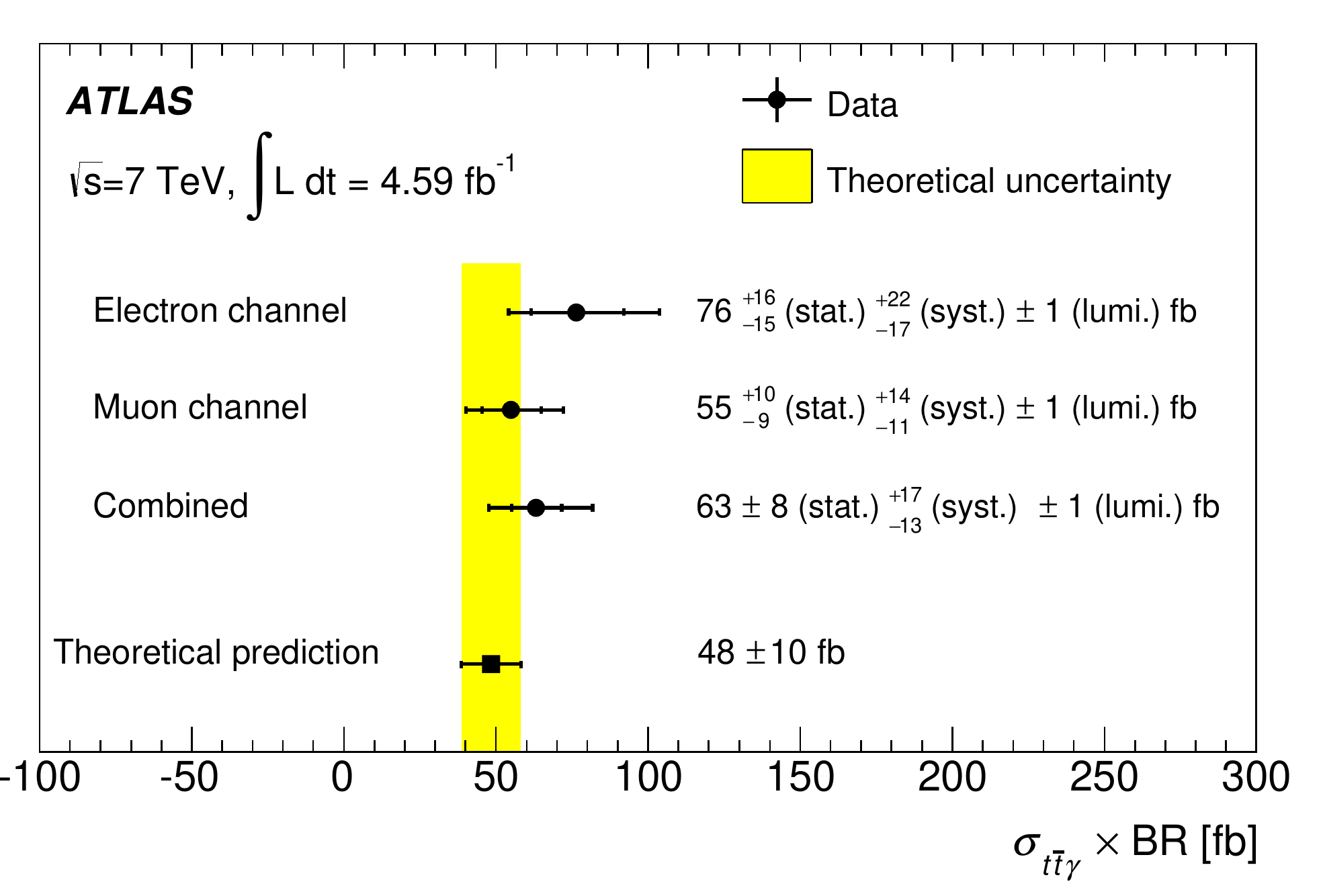}{0.51}{0}
\caption{Measurement of the \ttgam\ cross section at 7 \TeV~\cite{attgam}:  
results of the combined likelihood fit using the track-isolation ($\pt^{\mathrm{iso}}$)
distribution as the discriminating variable in the muon channel (left) and
summary of results compared to the theoretical prediction in the fiducial region
(right). \label{ttgamma}}
\end{figure}

The production cross section of top-quark pairs with additional photons is
sensitive to the \ttgam\ coupling and a measurement can be used to constrain
new physics, for instance with composite or excited top quarks.
With the full 7 \TeV\ dataset the ATLAS Collaboration reports
observation of this process with a significance of $5.3\,\sigma$~\cite{attgam}.
The measurement is performed in the single-lepton channel in a fiducial
region, requiring an identified photon with $E_T > 20 \GeV$, isolated
from jets and from leptons, in order to suppress the contribution
of photons not radiated from top quarks. The cross section is extracted from a
template fit to a variable that measures the track activity near the photon
candidate ($\pt^{\mathrm{iso}}$) and discriminates between signal photons and
neutral hadron decays or misidentified photons (Fig.~\ref{ttgamma}).  
With 362 selected events
the fiducial cross section is measured with an uncertainty 
of $\sim 30\%$, dominated by uncertainties on the jet energy scale and 
$b$-tagging efficiency.

At energies and luminosities available at the LHC the associated production 
of top-quark pairs with heavy vector bosons ($W$ or $Z$) become accessible. 
The production cross sections $\sigma_{\ttW}$ and $\sigma_{\ttZ}$ are 
simultaneously extracted in channels with two or more leptons, 
since the two processes are experimentally intertwined.
The interest in the measurement of the \ttZ\ process lies in the 
determination of the top-quark coupling to the $Z$ boson. The \ttW\ process
probes the proton structure and is a source of same-sign dilepton events, 
which is an important background in many searches.
A variety of new physics models can alter the prediction of 
$\sigma_{\ttW}$ and $\sigma_{\ttZ}$ and their effects are typically parameterised
by dimension-six operators in an effective field theory.

\begin{figure}[!h]\centering
\subfig{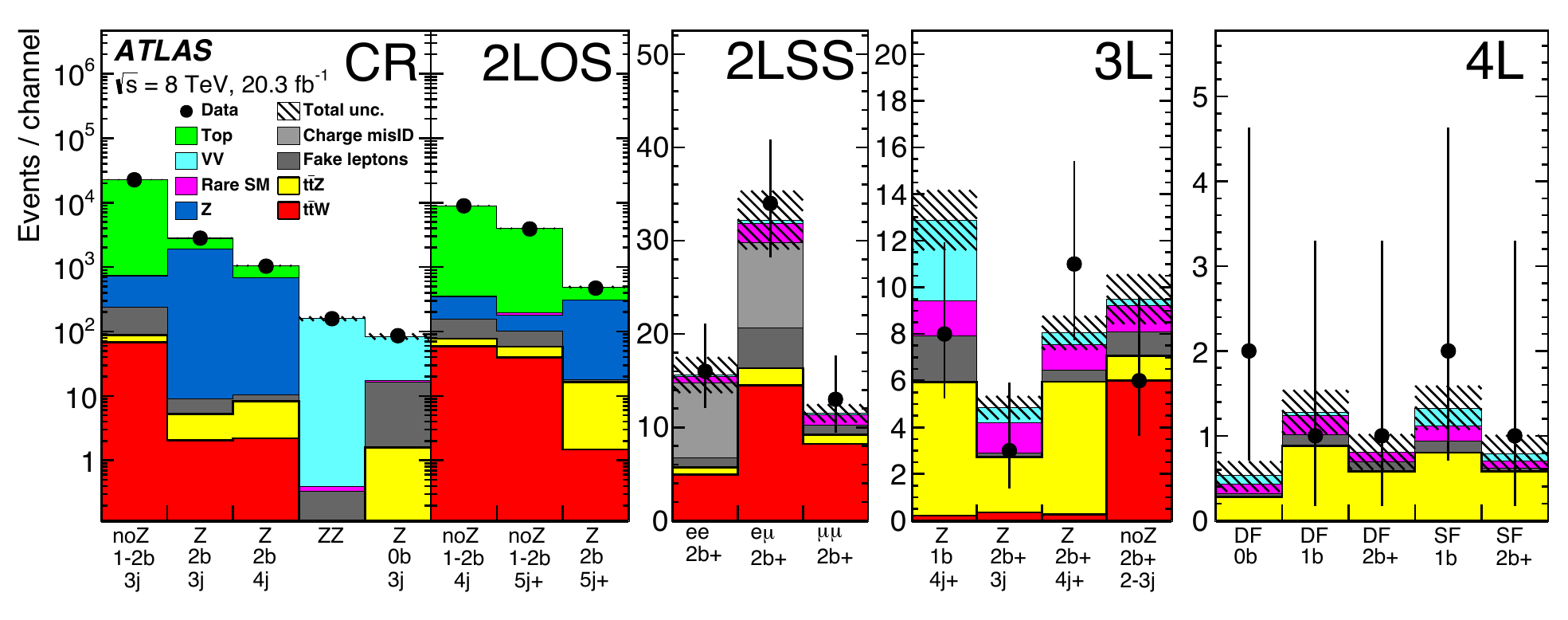}{0.99}{0}
\subfig{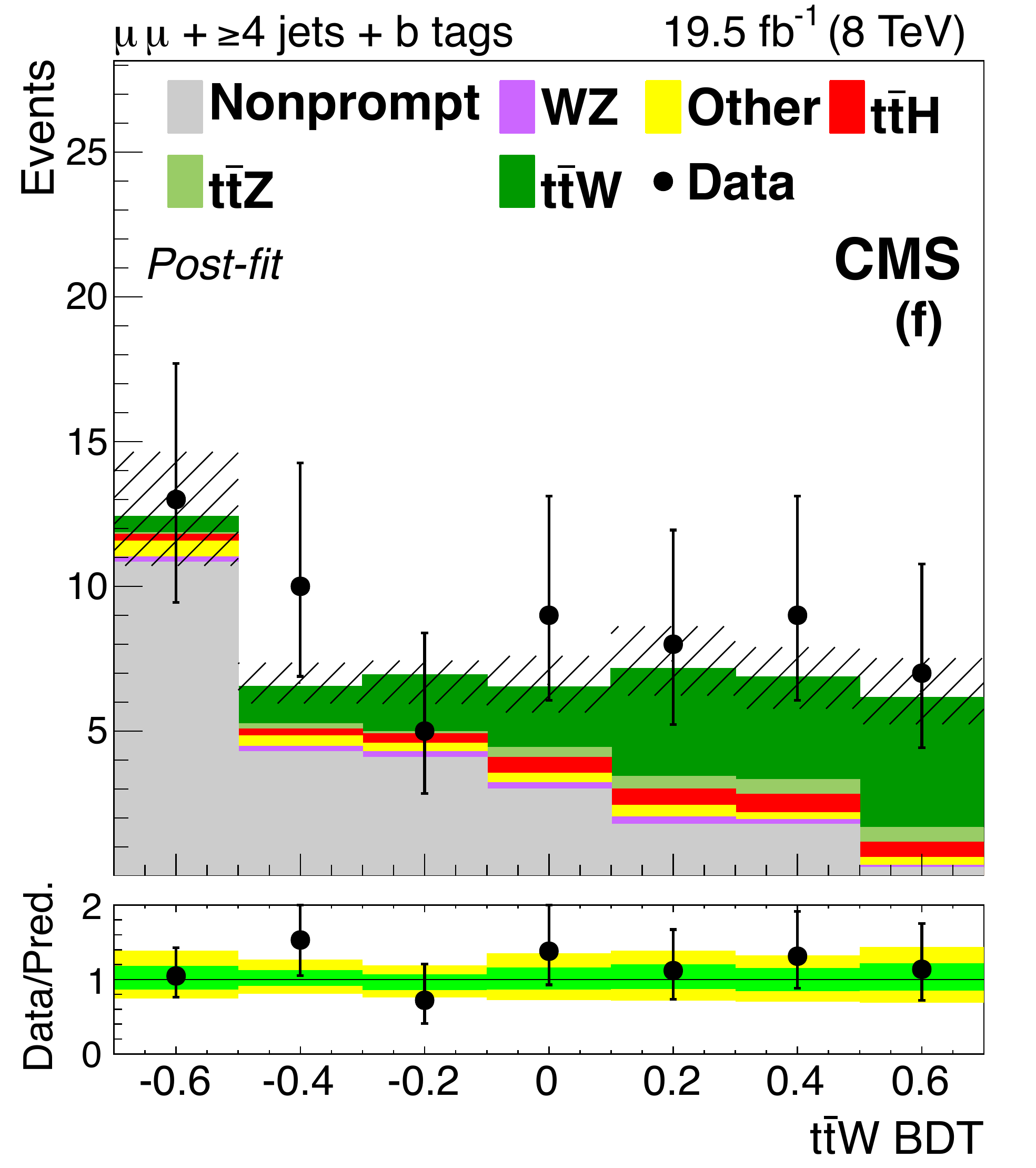}{0.36}{0}
\hspace{3ex}
\subfig{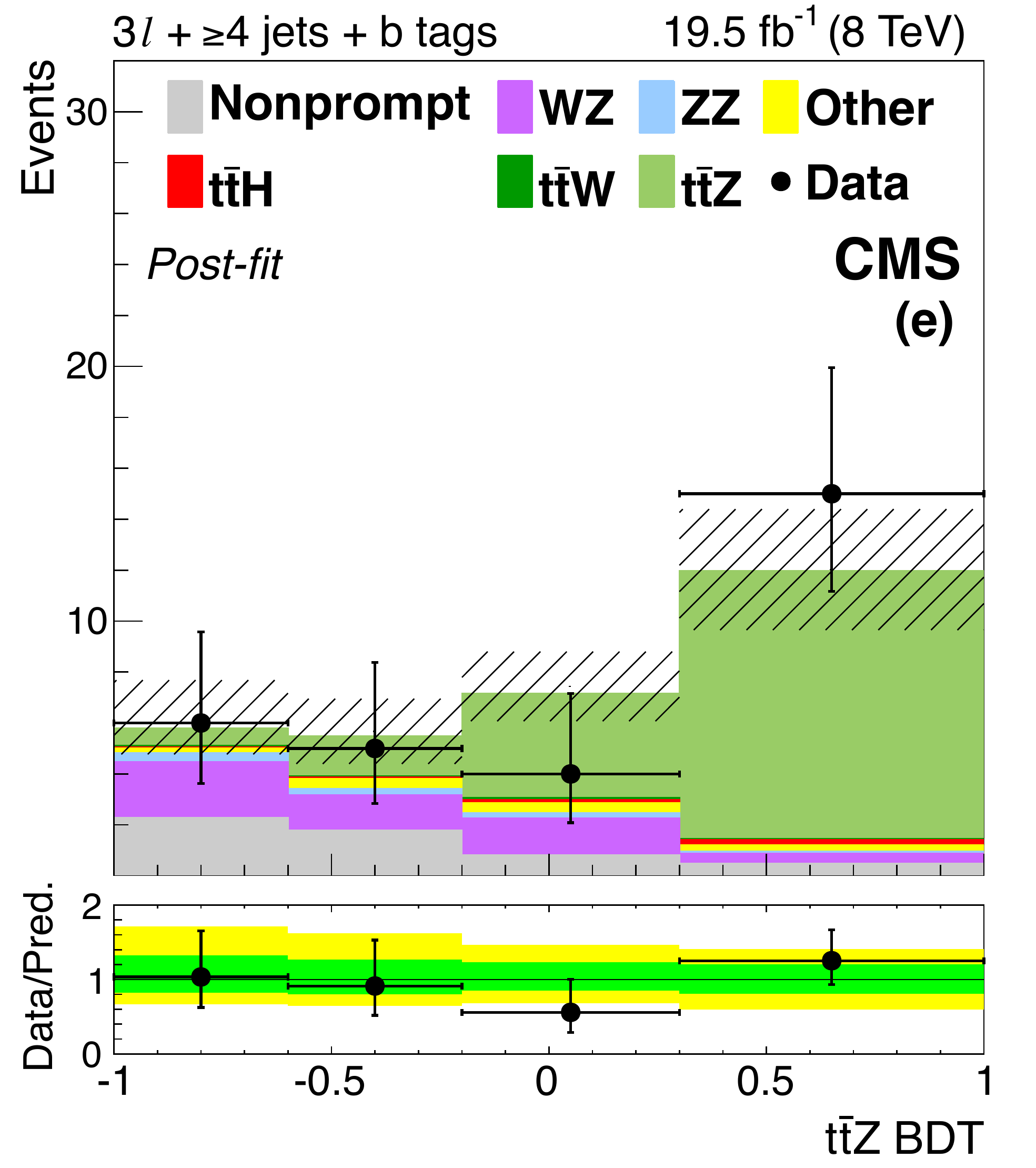}{0.36}{0}
\caption{Measurement of the \ttW\ and \ttZ\ cross sections at 8 \TeV.  Expected
yields after the fit compared to data in the four signal channels and five
control regions (CR) used to constrain the \ttbar, $Z$, $ZZ$ and $WZ$
backgrounds~\cite{attv} (top), and final discriminants for the \ttW BDT in the same-sign
$\mu\mu$ channel (bottom left) and the \ttZ BDT in the trilepton channel (bottom
right), both in events with at least four jets~\cite{cttv}.
\label{ttV1}}
\end{figure}

Using four signatures (opposite-sign dilepton, same-sign dilepton, trilepton,
and tetralepton) and performing a simultaneous maximum likelihood fit, ATLAS
extracts both cross sections with a significance of $5.0\,\sigma$ and
$4.2\,\sigma$, respectively, over the background-only hypothesis~\cite{attv}.  
In the opposite-sign dilepton channel the signal-to-background ratio is particularly
challenging and therefore control regions to determine the most important
backgrounds from data are included in the fit and a neural network is employed
in the signal regions. In the same-sign dilepton channel the contribution of
fake leptons and charge misidentification is carefully studied. The tri- and
tetralepton channels effectively measure the \ttZ\ contribution. Important
irreducible Standard Model backgrounds that produce three or four leptons are
diboson ($WZ$, $ZZ$) events, determined through fits in control regions, and
associated single top quark production ($tZ$, $WtZ$). The measured cross sections are
$\sigma_{\ttW} = 369^{+100}_{-91}$ fb and $\sigma_{\ttZ} = 176^{+58}_{-52}$ fb,
with uncertainties dominated by the statistical component (Fig.~\ref{ttV1} top).

\begin{figure}[!h]\centering
\subfig{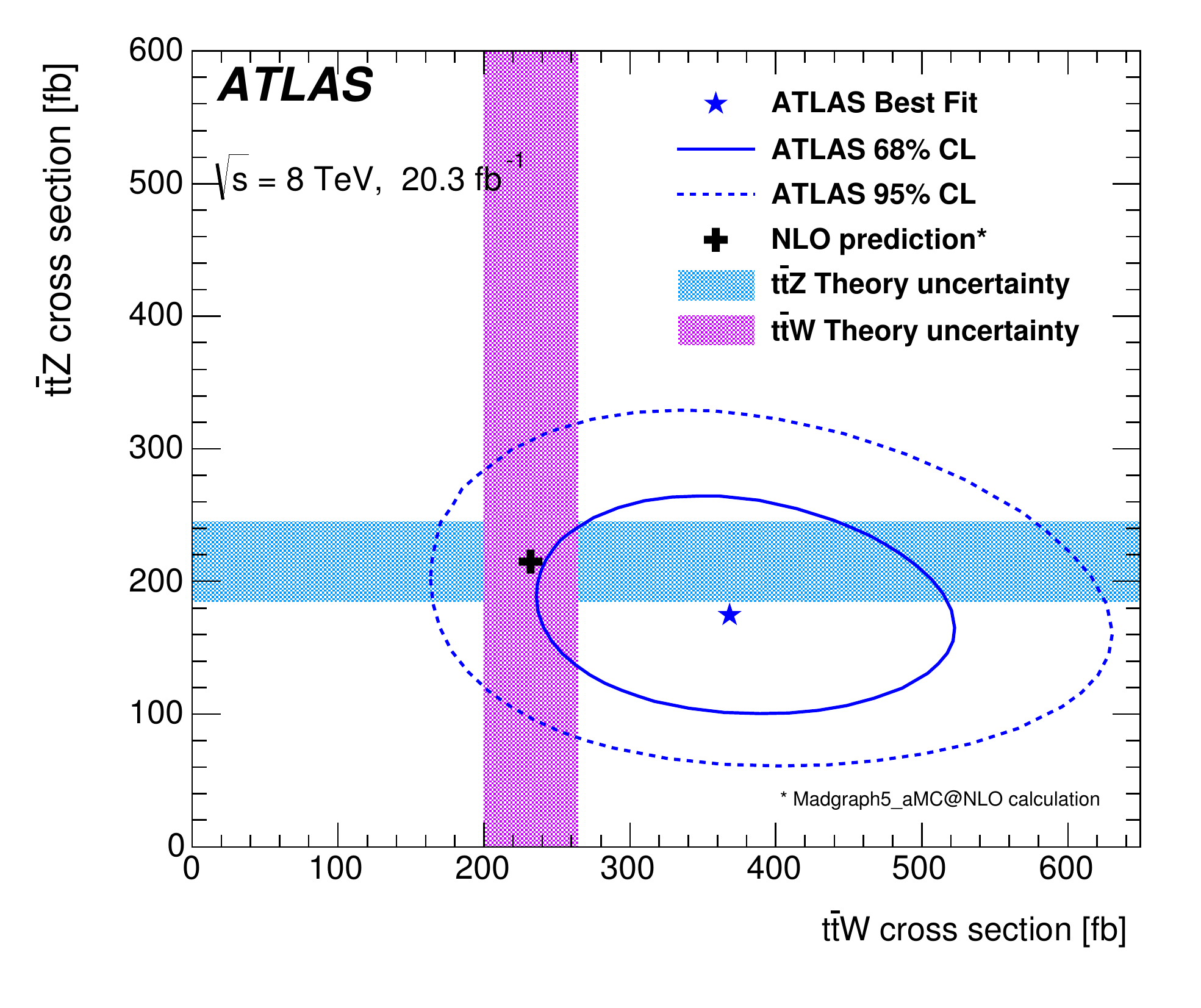}{0.47}{0}
\hspace{2ex}
\subfig{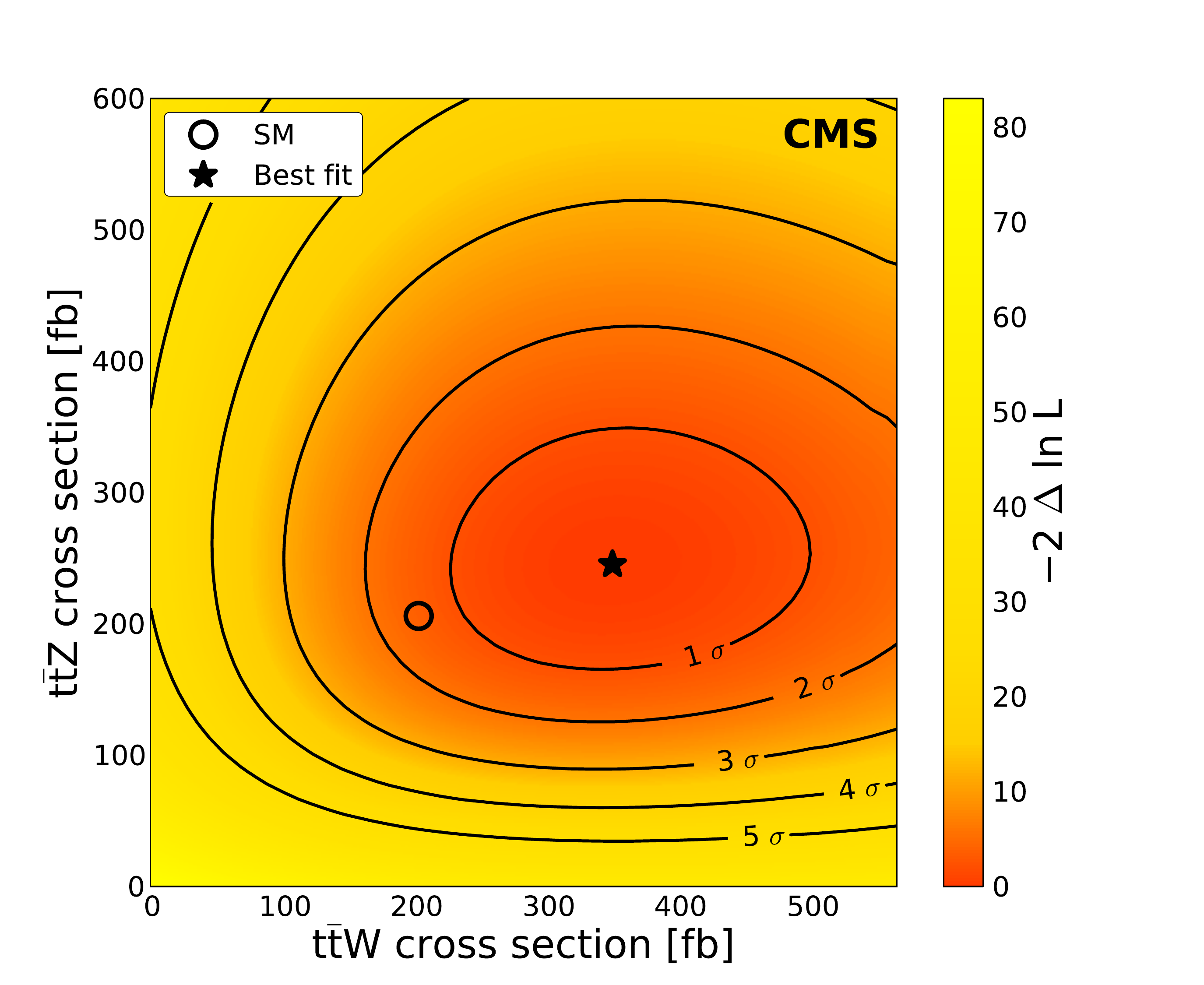}{0.48}{7}
\caption{Result of the simultaneous fit to the \ttW\ and \ttZ\ cross sections along
with uncertainty contours compared to theoretical predictions, for the ATLAS~\cite{attv}
 (left) and CMS~\cite{cttv} (right) measurements.
\label{ttV2}}
\end{figure}

CMS determines the cross sections of these processes, as well, 
with significances of $4.8\,\sigma$
and $6.4\,\sigma$, respectively~\cite{cttv}, using the same final states. The
analysis strategy aims at reducing the statistical uncertainty by lowering the
requirements on the reconstructed objects quality, at the expense of larger
systematic uncertainties. A full reconstruction of pre-selected events is
attempted, by matching the reconstructed objects in the detector to the decaying
$W$ and $Z$ bosons, and to the top quark. A linear discriminant helps to
determine the best permutation in matching jets and leptons. Signal is separated
from background by means of several BDTs, one in each channel and jet
multiplicity, which are trained with this discriminant and other kinematic
quantities (Fig.~\ref{ttV1} bottom). 
The measured cross sections are $\sigma_{\ttW} = 382^{+117}_{-102}$
fb and $\sigma_{\ttZ} = 242^{+65}_{-55}$ fb.
The result is then used to place constraints on the axial and vector components
of the \ttZ\ coupling and on dimension-six operators in an effective field 
theory framework.

\begin{figure}[!h]\centering
\subfig{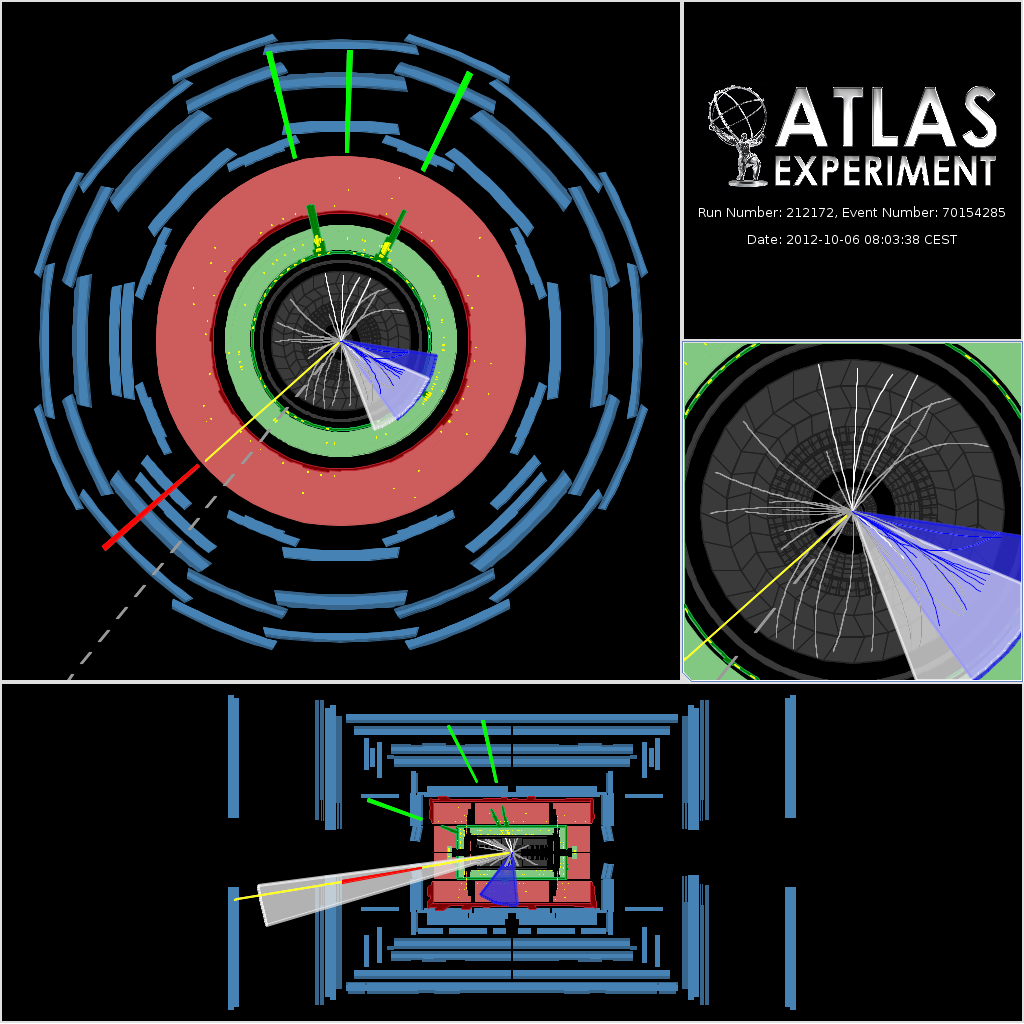}{0.8}{0}
\caption{
A display of a \ttZ\ candidate in the tetralepton channel~\cite{attv}. 
The event features a 
pair of electrons (green) with $m_{ee} = 93 \GeV$, and additionally a muon (red) 
and an extra electron, as well as two jets (blue and white cones), both of which 
are $b$-tagged. The missing transverse energy ($\MET = 57 \GeV$) is 
represented by the grey dashed line. \label{ttVevent}}
\end{figure}

Top-quark pairs are often produced with additional energetic jets. 
The measurement of such jet multiplicities provides an important test of the 
QCD predictions at NLO. Recently, the production of additional $b$-quarks has
been studied. The $\ttbar \bbbar$ final state is an irreducible non-resonant 
background to the \ttH process and difficult to model in simulation because of 
ambiguities in the matching to the parton shower. 

\begin{figure}[!h]\centering
\subfig{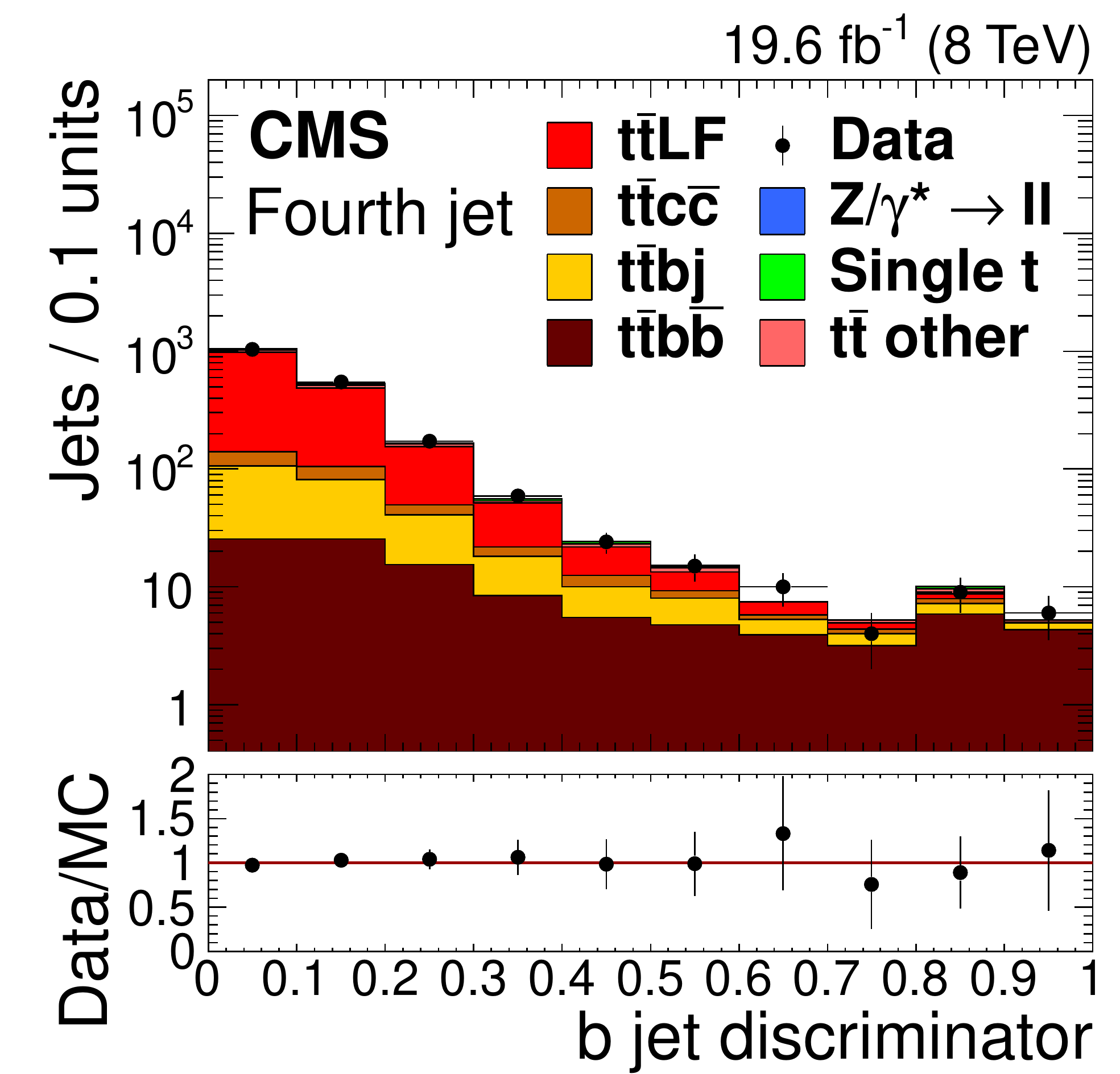}{0.3636}{0}
\hspace{5ex}
\subfig{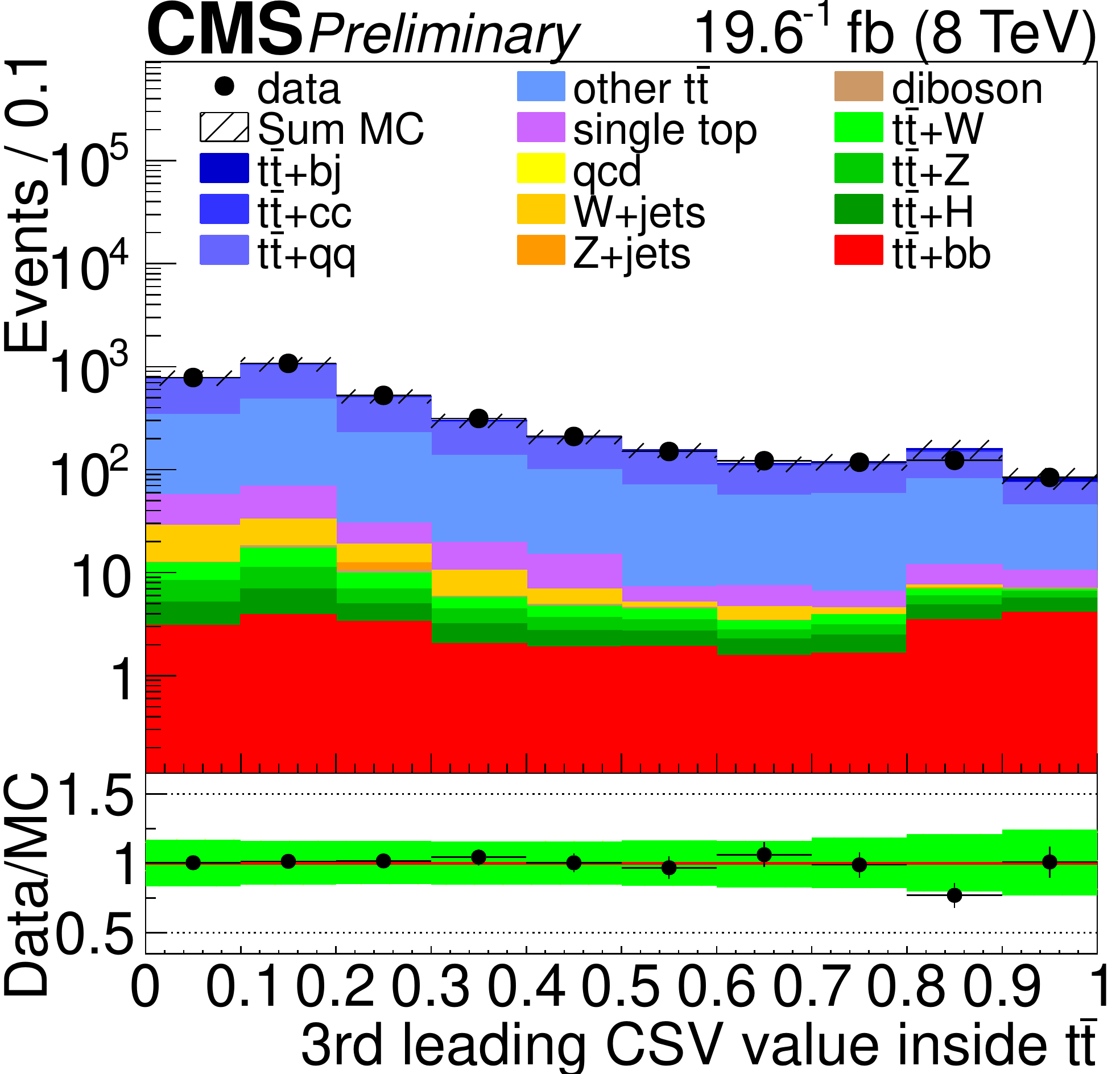}{0.3455}{0}
\subfig{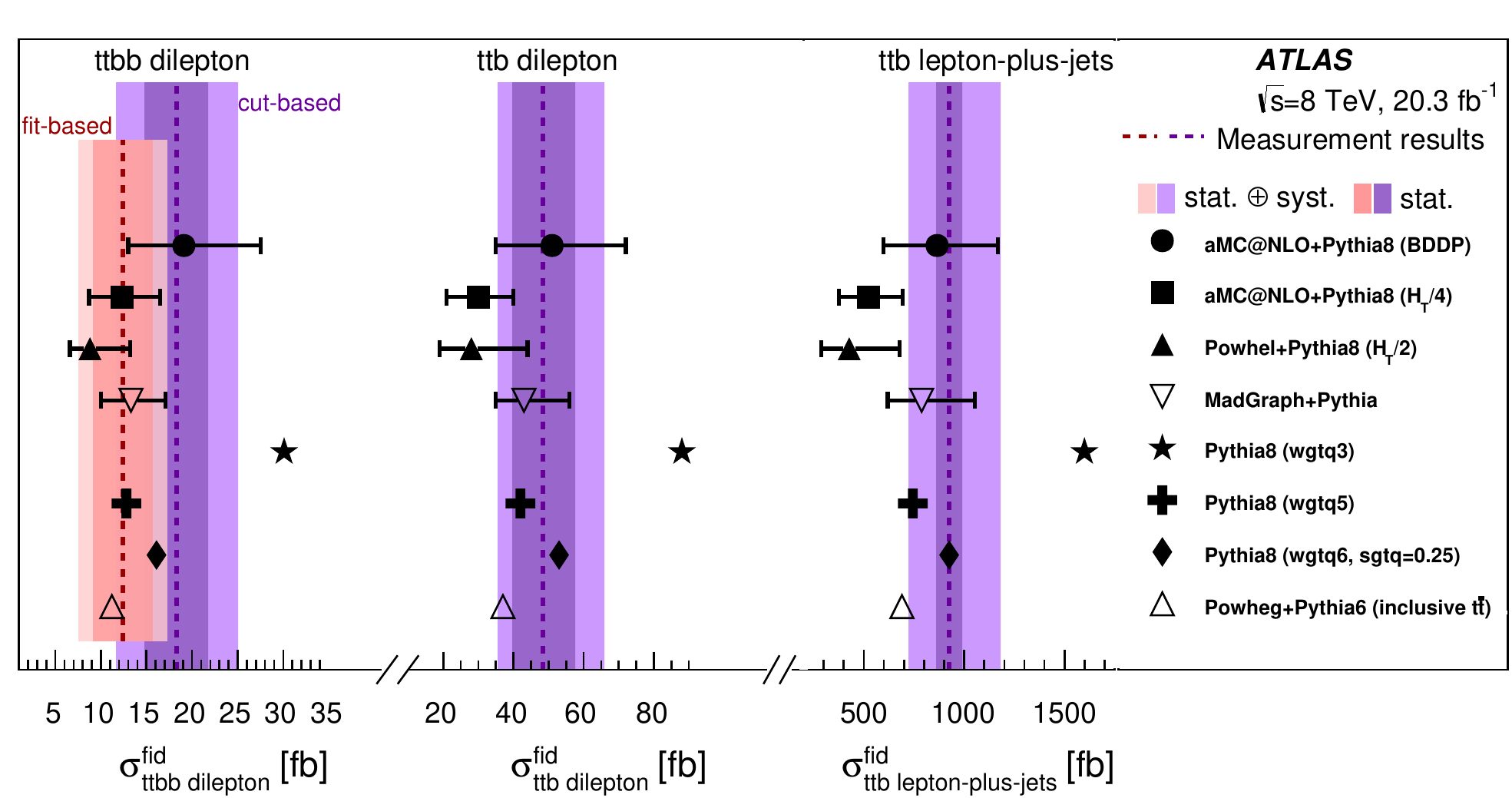}{0.616}{0}
\hspace{2ex}
\subfig{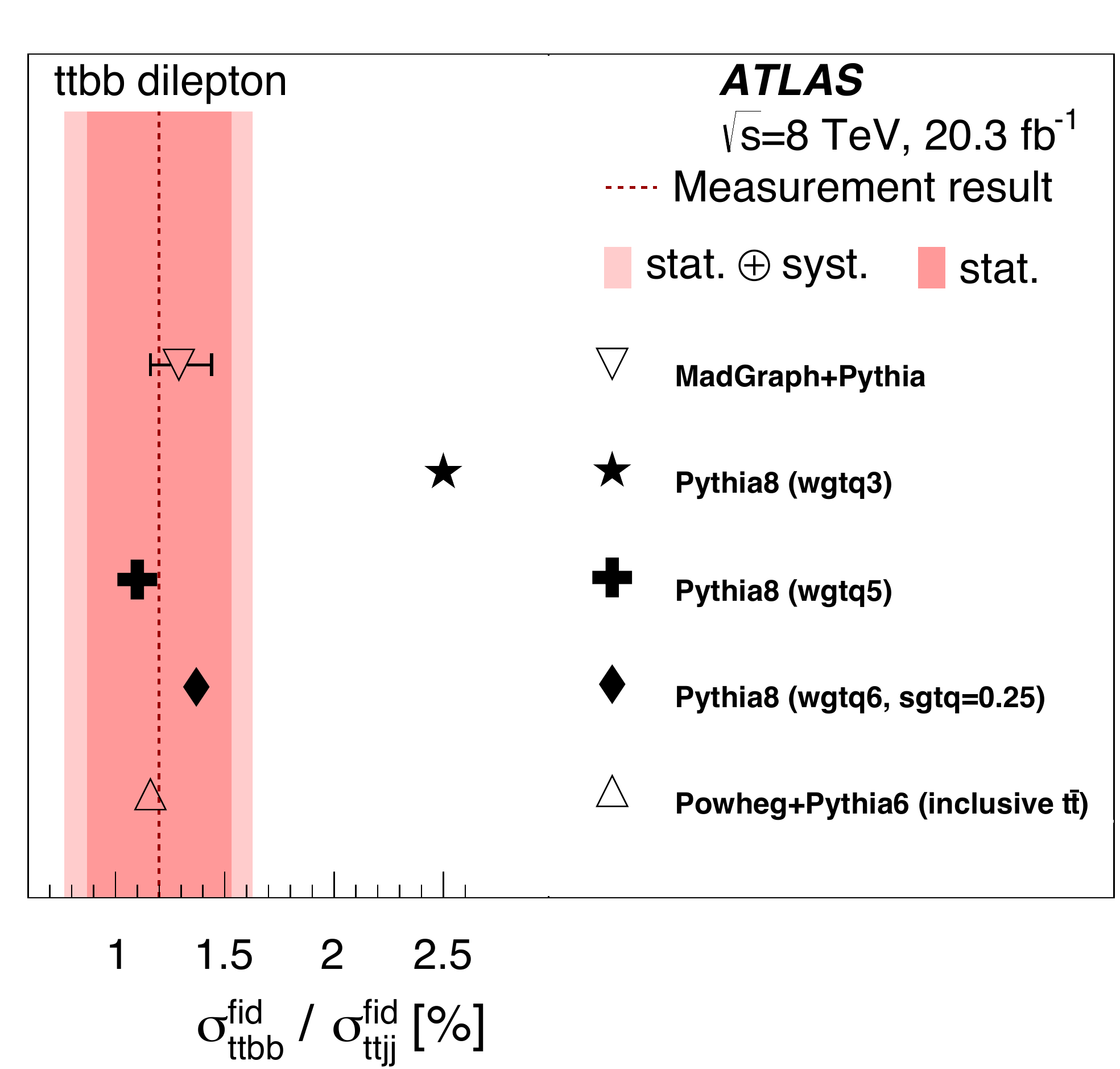}{0.340}{0}
\caption{Measurements of \ttbar with heavy flavour quarks at 8 \TeV.
Distribution of the $b$-jet discriminator for the fourth jet in decreasing order of
$b$-tagging discriminator value in the dilepton channel~\cite{ctthf} (top left), pre-fit
distribution of the $b$-tag discriminant of the third jet in the $\mu$+jets
channel for events with five jets~\cite{ctthf2} (top right), comparison of the measured 
cross sections in the three fiducial regions with theoretical predictions~\cite{atthf}
 (bottom left) and comparison of the measured ratio \RHF\ with different 
generator settings after subtraction of electroweak contributions~\cite{atthf}
 (bottom right).  \label{ttbb}}
\end{figure}

CMS measures the total cross section $\sigma_{\ttbar \bbbar}$ and the quantity
$\RHF = \frac{\sigma_{\ttbar \bbbar}}{\sigma_{\ttbar\!j\!j}}$ in the
dilepton~\cite{ctthf} and single-lepton channels~\cite{ctthf2}.  In the first
case, events with two well identified $b$-jets and at least two additional jets
are required. The $b$-tagging algorithm discriminator of the third and fourth
jet are used to separate $\ttbar \bbbar$ events from the background, including
$\ttbar\!j\!j$, with a template fit (Fig.~\ref{ttbb}).  
For the measurement in the single-lepton
channel the jets from the top-quark decay are identified using a constrained
kinematic fit and multivariate classifiers in different categories split by the
jet multiplicity.  Results range in $\RHF = 0.012 - 0.022$, depending on the
phase space and definition considered (particle or parton level) with
uncertainties of $0.004 - 0.006$, and are in general in good agreement with
predictions.

ATLAS performs four measurements of heavy flavour production in top-quark pair
events~\cite{atthf} in a fiducial volume: a fit-based and a cut-based
measurement of $\sigma_{\ttbar \bbbar}$ in the dilepton channel, and cut-based
measurements of $\sigma_{\ttbar b}$ in the dilepton and in the single-lepton
channels.  The ratio \RHF is determined to be $0.013 \pm 0.004$ with comparable
systematic and statistical uncertainties. The cut-based analysis uses very tight
selection criteria, including the requirement of four $b$-tagged jets,
relies on simulation for the background determination and features a
high signal-to-background ratio.  A looser selection is applied in the second
analysis where the signal is extracted from a fit to the multivariate $b$-jet
identification discriminant.  More events are produced in the single-lepton
channel, however this channel is affected by additional backgrounds, where a
$W$ boson can produce a $c$ quark.  The measurements are also presented after
subtracting the expected contributions from electroweak processes (\ttW, \ttZ
and \ttH) in order to allow for comparison with NLO QCD theory predictions.
They are then compared to predictions using different $g\to\bbbar$ splitting in
the parton shower. The most extreme \textsc{Pythia 8} model is disfavoured by the
measurements (Fig.~\ref{ttbb}).

The production of four top quarks is a rare process in the Standard Model,
proceeding via gluon-gluon fusion or quark-antiquark annihilation and is
predicted to have a total cross section of $\sim 1$ fb at $8 \TeV$.
The production can be significantly enhanced in BSM physics models and
therefore ATLAS and CMS searched for evidence of production of this process.
In a dedicated search~\cite{ctttt} CMS selects
events with one lepton, at least six jets and two $b$-tagged jets with 
large $H_{\mathrm{T}}$.
Kinematic reconstruction techniques and multivariate discriminants are employed
to place an upper limit on the production cross section.
ATLAS analyses events with a pair of leptons with the same charge with at least 
one $b$-tagged jet and large $H_{\mathrm{T}}$~\cite{atttt} to search for 
enhancements in the four-top production or evidence of new physics.
A search for vector-like quarks in the single-lepton channel by 
ATLAS~\cite{atttt2} is reinterpreted and used to set the best limit, 
$\sigma_{\ttbar \ttbar} < 23$ fb.

\begin{figure}[!h]\centering
\subfig{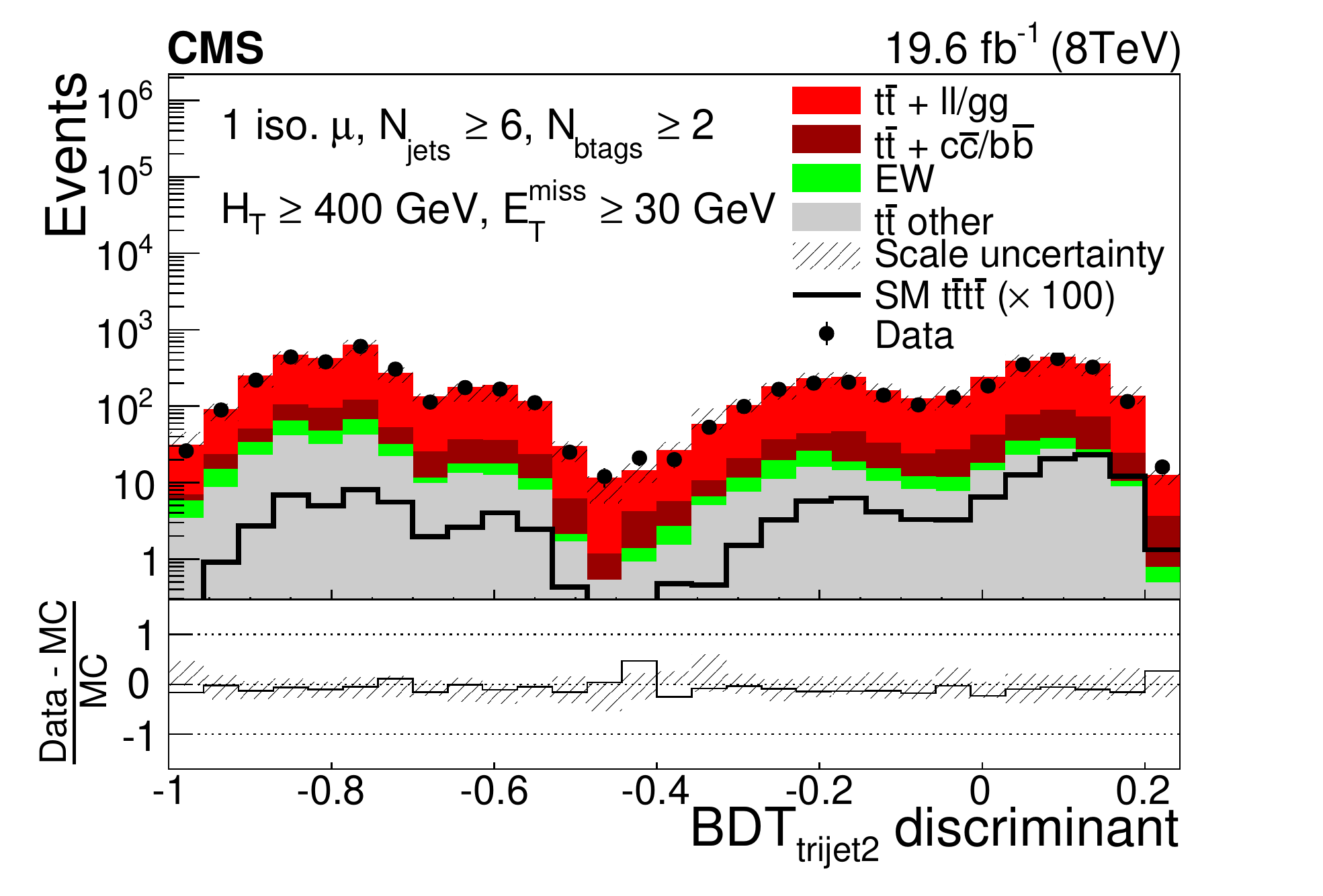}{0.65}{0}
\caption{Search for $\ttbar \ttbar$ production: Distribution of the BDT discriminant
aiming at identifying a second hadronically decaying top quark in the single-lepton
channel~\cite{ctttt}. \label{fourtops}}
\end{figure}

\section{Summary and conclusion}

Top-quark production measurements provide stringent tests of perturbative QCD
and have reached a good level of maturity with the large data samples available.
The inclusive top-quark pair production is
measured at centre-of-mass energies of $1.96$, $7$ and $8$ \TeV, with an uncertainty of
4--5\%, a better accuracy than the calculations at NNLO+NNLL.  First
measurements of the inclusive cross section are also available in the forward
region and at $13$ \TeV, albeit with larger uncertainties.  Differential
measurements allow to test in more depth the validity of the calculations and
the simulation programs that are used at hadron colliders. Results are reported
after unfolding for detector effects, extrapolating to either the parton or
particle level. The high-$\pt$ region is also explored, where top-quark decay
products start to merge and so-called boosted techniques improve the
reconstruction efficiency. In general, data and prediction match well, with some
discrepancies that might arise from the limited accuracy in the calculations
used in the simulation (NLO).  Single-top quark production has now been
observed in the $t$- and $s$-channels separately at Tevatron, while at the LHC
the $Wt$ channel has been observed in a combination and the abundant $t$-channel 
allows for differential measurements. The CKM matrix element \Vtb is determined with
an uncertainty of $4\%$ and is compatible with unity.  Associated production of
top-quark pairs with additional bosons or quarks have smaller production
cross sections and start to become accessible with the Run-1 dataset of LHC.
Production of \ttgam, \ttW and \ttZ have all been observed, while
upper limits are placed on $\ttbar\ttbar$.  Understanding additional heavy-flavour 
production is important, for instance as a background to \ttH, and is therefore matter of
detailed study.  With the upcoming Run-2 dataset, to be collected at $13 \TeV$ in
the next few years, it will be possible to challenge current calculations and
simulations at a next level of precision. 

\section{Acknowledgements}
The work of the author is currently funded by the European Research Council under 
the European Union's Seventh Framework Programme ERC Grant Agreement n.~617185.


\begin{thebibliography}{99}

\bibitem{melnikov} K.~Melnikov, \emph{Theoretical Results on Top Quark
Physics}, talk at this conference,
https://indico.cern.ch/event/325831/session/10/contribution/30.

\bibitem{properties} A.~Meyer, \emph{Determination of Top Quark Properties},
talk at this conference,
https://indico.cern.ch/event/325831/session/10/contribution/32.

\bibitem{tevcomb} T.~Aaltonen {\em et al.} (\coll{CDF and D0}s), \emph{Combination of measurements of the
top-quark pair production cross section from the Tevatron Collider}, 
\PRD\ 89 (2014) 072001.

\bibitem{d0xs} \coll{D0}, \emph{Measurement of the inclusive \ttbar production
cross section in \ppbar collisions at $\sqrt{s}=1.96$ \TeV}, D0 Note 6453-CONF,
www-d0.fnal.gov/Run2Physics/WWW/results/prelim/TOP/T106.

\bibitem{alt} \coll{ATLAS}, \emph{Measurements of the top quark branching ratios
into channels with leptons and quarks with the ATLAS detector}, 
\PRD\ 92 (2015) 072005.

\bibitem{alj} \coll{ATLAS}, \emph{Measurement of the top pair production cross
section in 8 \TeV\ proton-proton collisions using kinematic information in the
lepton+jets final state with ATLAS}, 
\PRD\ 91 (2015) 112013.

\bibitem{combtt} \coll{ATLAS and CMS}s, \emph{Combination of ATLAS and CMS top
quark pair cross section measurements in the $e\mu$ final state using
proton-proton collisions at $\sqrt{s} = 8$ \TeV}, \conf{2014}{054}{1319552},
\pas{14}{016}{1319376}.

\bibitem{toplhcwg} \coll{ATLAS and CMS}s, \emph{TOPLHCWG summary plots}, 
twiki.cern.ch/twiki/bin/view/LHCPhysics/TopLHCWGSummaryPlots.

\bibitem{lhcb} \coll{LHCb}, R.~Aaij {\em et al.}, \emph{First observation of top quark production in
the forward region}, 
\PRL\ 115 (2015) 112001.

\bibitem{beate} B.~Heinemann, \emph{ATLAS Results from Run2}, talk at this
conference, https://indico.cern.ch/event/325831/session/0/contribution/8.

\bibitem{luca} L.~Malgeri, \emph{CMS Results from Run2}, talk at this
conference, \mbox{https://indico.cern.ch/event/325831/session/0/contribution/9}.

\bibitem{a13} \coll{ATLAS}, \emph{Measurement of the \ttbar production
cross-section in $pp$ collisions at $\sqrt{s}=13$ \TeV\ using $e\mu$ events
with $b$-tagged jets}, \conf{2015}{033}{1385192}.

\bibitem{c13} \coll{CMS}, \emph{Measurement of the top quark pair production
cross section in proton-proton collisions at $\sqrt{s}=13$ \TeV}, 
\pas{15}{003}{1388254}.

\bibitem{cdiff8} \coll{CMS}, \emph{Measurement of the differential cross
section for top quark pair production in $pp$ collisions at $\sqrt{s} = 8$
\TeV}, 
\subm{\EPJC}, \arxiv{1505.04480}{ex}.

\bibitem{pseudo} \coll{ATLAS}, \emph{Differential top-antitop cross-section
measurements as a function of observables constructed from final-state
particles using pp collisions at $\sqrt{s}=7$ \TeV\ in the ATLAS detector}, 
JHEP 06 (2015) 100.

\bibitem{aboo} \coll{ATLAS}, \emph{Measurement of the differential
cross-section of highly boosted top quarks as a function of their transverse
momentum in $\sqrt{s} = 8$ \TeV\ proton-proton collisions using the ATLAS
detector},
\subm{PRD}, \arxiv{1510.03818}{ex}.

\bibitem{cboo} \coll{CMS}, \emph{Measurement of the differential \ttbar\
production cross section for high-\pt\ top quarks in $e/\mu$+jets final states
at $\sqrt{s} = 8$ \TeV}, \pas{14}{012}{1388555}.

\bibitem{tevst} T.~Aaltonen {\em et al.} (\coll{CDF and D0}s), \emph{Tevatron combination of
single-top-quark cross sections and determination of the magnitude of the
Cabibbo-Kobayashi-Maskawa matrix element $V_{tb}$}, 
\PRL\ 115 (2015) 152003.

\bibitem{ctchan} \coll{CMS}, \emph{Measurements of the differential cross
section of single top-quark production in the $t$ channel in proton-proton
collisions at $\sqrt{s} = 8$ \TeV}, 
\pas{14}{004}{1323200}.

\bibitem{aschan} \coll{ATLAS}, \emph{Search for $s$-channel single top-quark
production in proton-proton collisions at $\sqrt{s}=8$ \TeV\ with the ATLAS
detector}, 
\PLB\ 740 (2015) 118.

\bibitem{combWt} \coll{ATLAS and CMS}s, \emph{Combination of cross-section
measurements for associated production of a single top-quark and a $W$ boson at
$\sqrt{s}=8$ \TeV\ with the ATLAS and CMS experiments},
\conf{2014}{052}{1319379} and \pas{14}{009}{1319686}.

\bibitem{sinead} S.~Farrington, \emph{Results on the Standard Model Higgs boson}, 
talk at this conference, https://indico.cern.ch/event/325831/contribution/11.

\bibitem{attgam} \coll{ATLAS}, \emph{Observation of top-quark pair production
in association with a photon and measurement of the \ttgam production cross
section in $pp$ collisions at $\sqrt{s}=7$ \TeV using the ATLAS detector},
\PRD\ 91 (2015) 072007.

\bibitem{attv} \coll{ATLAS}, \emph{Measurement of the \ttW and \ttZ production
cross sections in $pp$ collisions at $\sqrt{s} = 8$ \TeV\ with the ATLAS
detector}, 
\accept{JHEP}, \arxiv{1509.05276}{ex}.

\bibitem{cttv} \coll{CMS}, \emph{Observation of top quark pairs produced in
association with a vector boson in $pp$ collisions at $\sqrt{s} = 8$ \TeV},
\subm{JHEP}, \arxiv{1510.01131}{ex}.

\bibitem{ctthf} \coll{CMS}, \emph{Measurement of the cross section ratio
$\sigma_{\ttbar \bbbar}/\sigma_{\ttbar\!j\!j}$ in $pp$ collisions at $\sqrt{s} = 8$
\TeV}, 
\PLB\ 746 (2015) 132.

\bibitem{ctthf2} \coll{CMS}, \emph{Measurement of the $\ttbar\bbbar$ cross
section and the ratio $\sigma(\ttbar\bbbar)/\sigma(\ttbar\!j\!j)$ in the
lepton+jets final state at 8 \TeV\ with the CMS detector},
\pas{13}{016}{1385669}.

\bibitem{atthf} \coll{ATLAS}, \emph{Measurements of fiducial cross-sections for
\ttbar production with one or two additional $b$-jets in $pp$ collisions at
$\sqrt{s}=8$ \TeV\ using the ATLAS detector}, 
\subm{\EPJC}, \arxiv{1508.06868}{ex}.

\bibitem{ctttt} \coll{CMS}, \emph{Search for standard model production of four
top quarks in the lepton + jets channel in $pp$ collisions at $\sqrt{s} = 8$
\TeV}, 
JHEP 11 (2014) 154.

\bibitem{atttt} \coll{ATLAS}, \emph{Analysis of events with $b$-jets and a pair
of leptons of the same charge in $pp$ collisions at $\sqrt{s} = 8$ \TeV\ with
the ATLAS detector}, 
JHEP 10 (2015) 150.

\bibitem{atttt2} \coll{ATLAS}, \emph{Search for production of vector-like quark
pairs and of four top quarks in the lepton-plus-jets final state in $pp$
collisions at $\sqrt{s} = 8$ \TeV\ with the ATLAS detector}, 
JHEP 08 (2015) 105.

\end{thebibliography}
\end{document}